\documentclass[11pt,a4paper]{article}
\usepackage[T1]{fontenc}
\usepackage[latin9]{inputenc}
\usepackage{bm}
\usepackage{amsmath}
\usepackage{graphicx}
\makeatletter
\usepackage[ngerman,english]{babel}
\newif\ifAMStwofonts
\AMStwofontstrue

\def\gsim{~\rlap{$>$}{\lower 1.0ex\hbox{$\sim$}}}

\def\simpropto{\lower.2ex\hbox{$\; \buildrel \propto \over \sim \;$}}
\def\ltsim{\lower.5ex\hbox{$\; \buildrel < \over \sim \;$}}
\def\gtsim{\lower.5ex\hbox{$\; \buildrel > \over \sim \;$}}
\def\ltsim{\lower.5ex\hbox{$\; \buildrel < \over \sim \;$}}
\def\gtsim{\lower.5ex\hbox{$\; \buildrel > \over \sim \;$}}

\def\pmb#1{\setbox0=\hbox{#1}%
\kern-.025em\copy0\kern-\wd0
\kern.05em\copy0\kern-\wd0
\kern-.025em\raise.0433em\box0}

\def\simlt{\lower.5ex\hbox{$\; \buildrel < \over \sim \;$}}
\def\simgt{\lower.5ex\hbox{$\; \buildrel > \over \sim \;$}}

\newcommand{\beq}{\begin{equation}}

\newcommand{\eeq}{\end{equation}}
\def\beqa{\begin{eqnarray}}
\def\eeqa{\end{eqnarray}}
\def\fixit#1{}

\usepackage{soul}

\usepackage{xcolor}
\usepackage{esint}
\usepackage{soul}
\setstcolor{red}

\newcommand{\pt}[1]{{\color{magenta}#1}}

\usepackage{tikz}
\usetikzlibrary{arrows.meta}
\tikzset{%
  >={Latex[width=2mm,length=2mm]},
            base/.style = {rectangle, rounded corners, draw=black,
                           minimum width=4cm, minimum height=1.5cm,
                           text centered, font=\sffamily},
  activityStarts/.style = {base, fill=blue!30},
       startstop/.style = {base, fill=red!30},
    activityRuns/.style = {base, fill=green!30},
         process/.style = {base, minimum width=2.5cm, fill=orange!15,
                           font=\sffamily},
}

\usepackage{bm}
\pdfoutput=1 % if your are submitting a pdflatex (i.e. if you have
\usepackage{jcappub}% for details on the use of the package, please
\usepackage[normalem]{ulem}

\title{A study of Dipolar Signal in distant Quasars {with various observables}}
\author[a]{Rahul Kothari,}
\author[b]{Mohit Panwar,}
\author[b]{Gurmeet Singh,}
\author[c]{Prabhakar Tiwari,}
\author[d]{Pankaj Jain}

\affiliation[a]{Department of Physics \& Astronomy, University of the Western Cape, Cape Town 7535, South Africa}
\affiliation[b]{Department of Physics, Indian Institute of Technology, Kanpur 208016, India}
\affiliation[c]{National Astronomical Observatories,
Chinese Academy of Science, Beijing, 100101, P.R.China}
\affiliation[d]{Department of Space Science \& Astronomy, Indian Institute of Technology, Kanpur 208016, India}

\emailAdd{quantummechanicskothari@gmail.com}
\abstract{We study the signal of anisotropy in AGNs/quasars of CatWISE2020 catalogue using different observables. It has been reported earlier that this data shows a strong signal of dipole anisotropy in the source number counts. We test this claim using two independent data analysis procedures and find our number count dipole consistent with the earlier results. In addition to number counts, we test for the anisotropy signal in two other observables --- mean spectral index $\bar{\alpha}$ and mean flux density $\bar{B}$. We find a dipole signal of considerable strength both in the mean spectral index and the mean flux density. The dipole in mean flux density points towards the galactic center and becomes very weak after imposing a flux cut to remove sources with flux greater than 1 mJy. This can be attributed to the presence of some bright sources. The signal in mean spectral index, however, is relatively stable as a function of both flux and galactic cuts. The dipole in this observable points roughly opposite to the 
galactic center and hence most likely arises due to galactic bias. Hence, the signal in both the mean spectral index and mean flux density appears to be consistent with isotropy. } 

\makeatother

\begin{document}
\maketitle \flushbottom

\section{Introduction\label{sec:Introduction}}
The Standard Model of Cosmology{,} aka $\Lambda$CDM{,} summarizes our current understanding of the Universe. {One of the underlying assumptions of the model is} 
the Cosmological Principle \citep{Bengaly:2018,Nadolny:2021hti,Kim:2021osl,Bengaly:2021wgc,Lahav:2000ra,Aluri:2022hzs}, according to which the Universe is statistically isotropic and homogeneous at sufficiently large length scales \citep{Shamik:2016,Bengaly:2019zhr}. The precise value of this distance is still not clear but is expected to be 100 Mpc or larger \citep{Kim:2021osl}. Additionally, the principle is expected to hold in a special frame \citep{Weinberg:2008,Horstmann:2021jjg} called the Cosmic Rest Frame (CRF henceforth). In this frame, all cosmological observables are expected to be statistically isotropic. 
{Because the solar system is moving with respect to this frame, several cosmological observables, including the Large Scale Structure (LSS) and the Cosmic Microwave Background (CMB), are expected to exhibit dipole anisotropy in this frame due to Doppler shift and aberration effect\st{s}
\citep{Kogut:1993,Hinshaw:2009}}. The dipole in the CMB has been measured very accurately and has been used to predict our velocity with respect to CRF. The LSS {dipole} has also been observed, but its magnitude does not appear to agree with the velocity predicted by the CMB dipole (see \cite{Darling:2022jxt} for a contrary view), indicating a {potential} departure from the cosmological principle \citep{Singal:2011,Gibelyou:2012,Rubart:2013,Tiwari:2014ni,Secrest:2020CPQ,Secrest:2022uvx}. In this paper, we revisit the dipole in number counts using {the} WISE data catalogue.  {In addition to this}, we use two other observables -- (a) mean spectral index $\bar{\alpha}$ and (b) mean flux density $\bar{B}$. Our analysis is based on the extraction of first three multipoles from the data. 
Inclusion of quadrupole  accounts for the leakage of dipole power into its neighbouring multipoles. The power beyond quadrupole is found to be negligible and hence we have neglected any multipoles beyond quadrupole in our analysis.

The dipole in LSS has been measured using radio surveys  \citep{Singal:2011,Gibelyou:2012,Rubart:2013,Tiwari:2014ni}, as well as quasars, observed at Infrared frequencies using the catWISE2020 catalogue \citep{Secrest:2020CPQ}. In both cases,
the observed dipole \textcolor{blue}{in number counts} has a much higher amplitude in comparison to the local motion based
CMB prediction. Additionally, the direction is close to the CMB dipole \citep{Blake:2002ac,Singal:2011,Gibelyou:2012,Rubart:2013}. A similar behaviour is seen in the radio polarized flux \citep{Tiwari:2015np}. The dipole signal \textcolor{blue}{in number counts,} observed by \cite{Secrest:2020CPQ} appears to show a strong, $4.9\sigma$ excess, in comparison to the CMB based prediction. Thus{,} the dipole amplitude is found to be roughly two
times higher as compared to the CMB dipole. {The direction of the
 dipole is found to be {(}$l=238.2^{\circ},b=28.8^{\circ}${)} in galactic coordinates. }

There {are} also many other observations which appear to show that the Universe
is not statistically isotropic even on very large distance scales{. For} example {Park et. al. (2016)} \cite{Park:2016xfp} perform isotropy and homogeneity test{s} on {the} SDSS Luminous Red Galaxy sample and find that even at $300h^{-1}$ Mpc, the {isotropy} doesn't seem to hold. In the case of CMB, a potential violation of isotropy  has been termed \textit{hemispherical power asymmetry} \citep{Eriksen:2004} which is basically the presence of different CMB powers in different hemispheres. It still persists in the data {at} around $3\sigma$ statistical significance \citep{2007ApJ...660L..81E,refId0, Planck_iso:2016, Planck_iso:2013, Hoftuft:2009rq}. {A detailed review of isotropy violations is given in Ref. \cite{Aluri:2022hzs}.} It is somewhat interesting that several of these observations indicate a preferred direction that is closely aligned  with the CMB dipole \citep{Ralston:2004}. These include the dipole in the radio polarization offset angles \citep{Jain:1998r} and the alignment of the CMB quadrupole and octopole \citep{2004PhRvD..69f3516D,Schwarz:2004}. The quasar optical polarizations show an alignment over very large distance scales \citep{Hutsemekers:1998,Jain:2003sg}. This alignment also has a tendency to maximize in the direction close to the CMB dipole \citep{Ralston:2004}. 

If the observed deviations from isotropy are indeed confirmed,
they would potentially require a major departure from the standard $\Lambda$CDM cosmology. There have been many theoretical attempts to explain these observations \citep{Jain:2002vx,Morales:2007rd,Urban:2009sw,Piotrovich:2009zz,Agarwal:2009ic,Poltis:2010yu,Hackmann:2010ir,Agarwal:2012,Tiwari:2016sp,Tiwari:2021si}. Here we mention one potential explanation that requires a minimal departure from the $\Lambda$CDM model
\citep{Gordon:2005,Erickcek:2008} and is based on the existence of
superhorizon modes \citep{Grishchuk:1978AZh, Grishchuk:1978SvA}.
The superhorizon modes have wavelengths much larger than the horizon size and hence do not have much effect on most cosmological observations \citep{Grishchuk:1978AZh, Grishchuk:1978SvA}. In order to explain the observed violations of isotropy, these modes are assumed to be aligned with one another beyond a certain length scale, i.e., their wave vectors $\mathbf{\hat{k}}$ point in the same direction.
In \citep{Shamik:2014,Das:2021}, the authors implemented this mechanism by assuming the existence of just one such mode and showed that the observed dipole in large scale structures
can be explained.
Such a mode also affects the CMB quadrupole and octopole \citep{Gordon:2005,Erickcek:2008} and can potentially explain their observed alignment \citep{2004PhRvD..69f3516D}. Remarkably, this simple model also explains the observed tension in the Hubble parameter \citep{Tiwari:2021ikr}.
It is rather interesting that this model may emerge from a pre-inflationary phase
%of the history of the Universe
\citep{Aluri:2012,Rath:2013}. The Universe need not
be isotropic and homogeneous before inflation and is expected to acquire this property
within the first inflationary efold\footnote{Explicit proofs exist in the case of {homogeneous comsologies like} Bianchi \citep{Collins:1972tf,Wald:1983}, {Kantowski Sachs \citep{Collins:1977fg,Turner:1986gj,Leon:2010pu, Fadragas:2013ina}, also some cases of inhomogeneous cosmologies \citep{Stein-Schabes:1986lic,Jensen:1986vs}}.}. Hence, we expect that at some sufficiently large
distance scales, the cosmic modes need not follow the cosmological principle. This distance scale
corresponds to the wavelength of the modes {that} left the horizon during {the} aforementioned early stage of
inflation. Hence{,} it is possible that these observed deviations from isotropy may be pointing
towards the physics of such an early stage of inflation. Irrespective of this relationship, this model
provides a simple and {viable explanation} %testable model 
for such observations. The model leads to
small anisotropy in many cosmological
observables, whose amplitude is expected to be of the order of the dipole in
large scale structures or smaller.

In this paper, we explore the dipole signal in {three}
observables using the catWISE2020 data. {We account for the leakage of dipole power into its adjacent multipole--the quadrupole, and extract the dipole in observables by simultaneously fitting first three multipoles. The contribution beyond quadrupole is relatively small and can be neglected. Hence, in our analysis, we don't consider the octopole and higher multipoles.} We revisit the signal in number counts using two methods different  from \cite{Secrest:2020CPQ} and 
{thus provide an independent validation of their findings.} 
The number counts acquire a dipole distribution due to our motion with respect to the CRF \cite{Ellis:1984}. The analysis in \cite{Secrest:2020CPQ} shows that the amplitude of this dipole is much higher than expected on  {purely} kinematic grounds. 
Given the significance of the effect claimed in \cite{Secrest:2020CPQ}, it is clearly important to determine its source. Assuming that the signal has a physical origin, it is very
likely that other observables may also show an anisotropic behaviour. 
A study of dipole in other observables can help in disentangling
the physical origin of the effect.
We identify two such observables which we study in detail in this paper.
These are:
\begin{enumerate}
\item The {mean} spectral index ($\bar{\alpha}$): The variation of flux density $S(\nu)$ ($\mathrm{Wm^{-2}{Hz}^{-1}}$) as a function of frequency $(\nu)$ %\st{of the cosmic sources  generally} 
follows a power law
 $$S(\nu) \propto \nu^{-\alpha}$$
Although flux density $S$ changes in two different reference frames, $\alpha$ doesn't. 
We define the mean spectral index $\bar{\alpha}$ as the sum of spectral indices in a given patch of sky divided by the total number of sources in that patch. In the standard Big Bang cosmology, $\bar{\alpha}$ would be distributed isotropically in the sky. It is not expected to get any contribution due to kinematic effects. {Hence any anisotropy in this parameter would  signal a non-standard cosmology.} %\st{It must be noted} 
{Notice} that this observable has been used in \cite{Secrest:2022uvx} for flux calibration.

\item The mean {flux density} ($\bar{B}$): It is obtained by
dividing the total flux with the number of sources in any region of the sky. %Thus, in each pixel, it is computed by dividing the total flux with the number of sources in that pixel. 
If we assume that the {integral} number counts {above some flux density $S$} show a power law distribution, i.e.  ${\mathrm{d}N\over \mathrm{d}\Omega} (>S) \propto S^{-x}$, 
 then $\bar{B}$ would be distributed isotropically \citep{Tiwari:2019TGSS}.
{Assuming a kinematic origin of dipole, this observable would get a non-zero contribution only if the number count distribution differs from a pure power law. Else, a dipole in this observable can arise only in a
 non-standard cosmology.}
\end{enumerate}
These two observables are also interesting since they are unaffected by the distribution of sources in the sky. 

The paper is structured in the following manner. In \S\ref{sec:catWiseDetail}, we give details of the catWISE2020 catalogue. %This also contains details pertaining to our map preparation method. 
After this, we explain our first method; based on a $\chi^2$ minimization, for multipole recovery from the masked sky, in \S\ref{sec:MultipoleRecovery}. {Our second method is based on an extension of \texttt{Healpy fit\_dipole} method. We apply this method to number counts $N$ and find consistency between the two methods.} %The other one is an extension of the \texttt{Healpy fit\_dipole} 
{The details of the second} method are discussed in Appendix \ref{sec:partialMethod} where we also estimate the errors in the quantities of interest by simulations. %{We have also explained our procedure with an example.} 
In \S\ref{sec:Results}, we discuss dipole anisotropy results for $N$, $\bar{\alpha}$ and $\bar{B}$ obtained using $\chi^2$ method. Further, we study dipole anisotropy in $\bar{\alpha}$ and $\bar{B}$, using both galactic and flux cuts. In $\bar{\alpha}$, we find a strong signal of dipole anisotropy  and the direction lies close to the galactic plane. In $\bar{B}$, the dipole signal is found to be strong to mild depending upon the imposed flux and galactic cuts. For the cases where the dipole signal is strong, the direction lies close to the galactic plane. 
We conclude in \S\ref{sec:Conclusion}. 

\section{The CatWISE catalogue \label{sec:catWiseDetail}}
\begin{figure}
	\centering
	\includegraphics[scale=0.35]{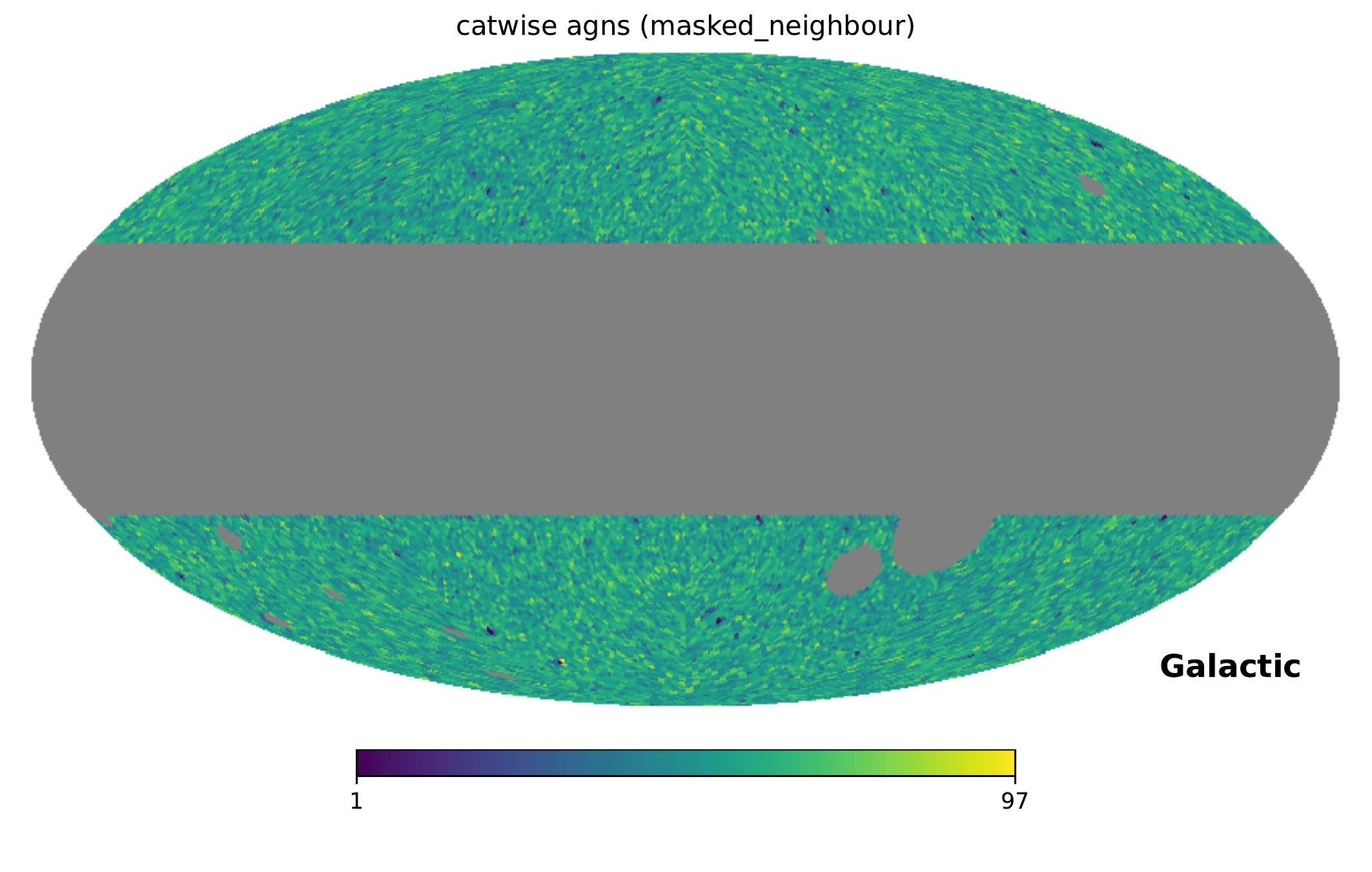}
	\includegraphics[scale=0.35]{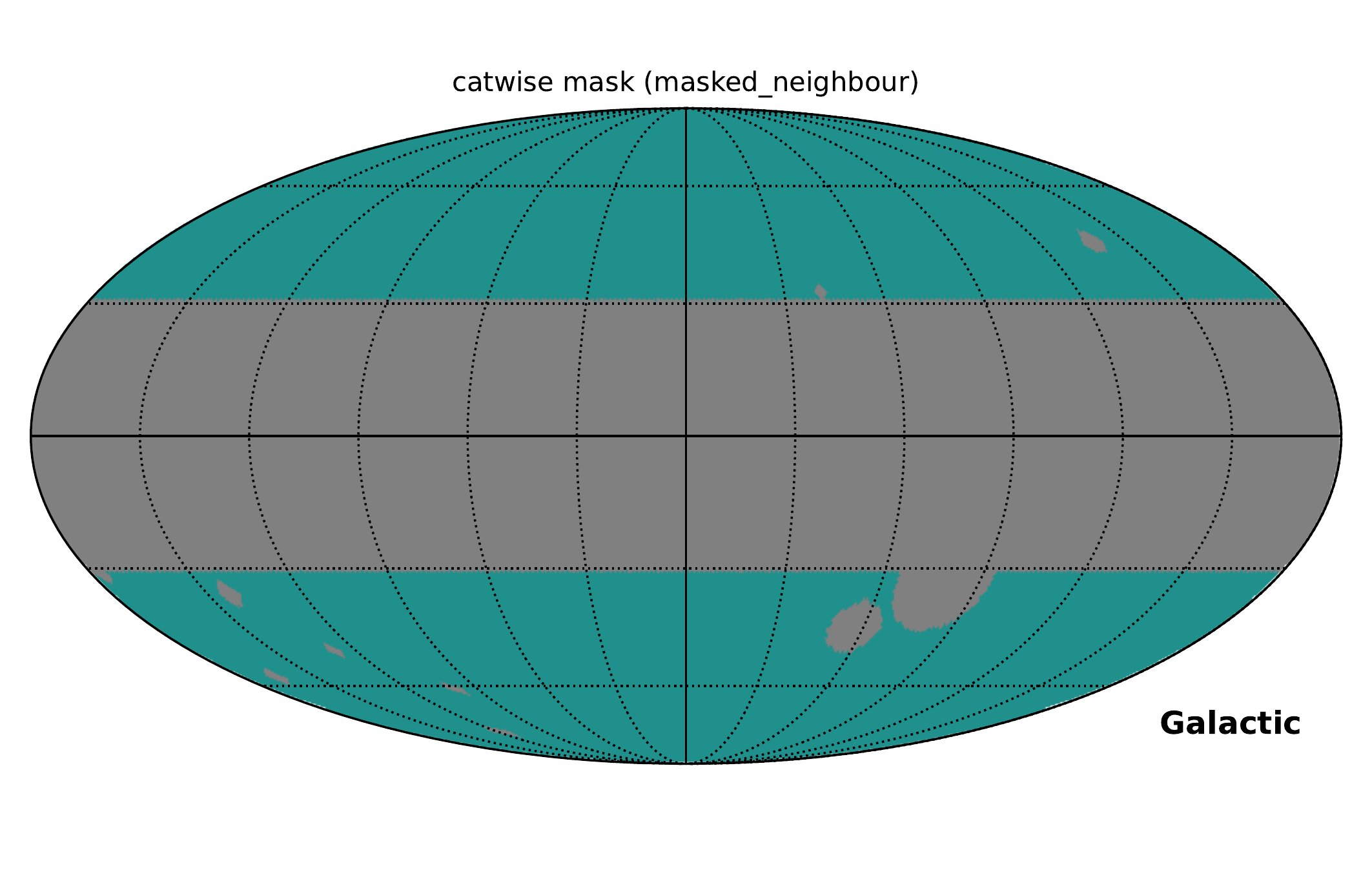}
	\caption{\textit{Left:} Number count map and  \textit{Right:} the corresponding mask map for catWISE2020 catalogue after following the masking procedure described in the text.}
	\label{fig:Data_Mask}
\end{figure}

The catWISE2020 catalogue is generated from the \textit{Wide-field Infrared Survey Explorer (WISE)} \citep{2010AJ....140.1868W} and \textit{NEOWISE} all-sky survey data at infrared wavelengths $3.4\mu$m and $4.6\mu$m in {the} W1 and W2 bands{,} respectively. 
%$(W1\: and\:W2\: band)$. 
It contains 1,890,715,640 sources and results from the catWISE Preliminary catalogue \citep{2020ApJS..247...69E}. The latter is generated from the data collected from 2010 to 2016, and two years additional data from the survey.
The catWISE2020 has 90\% completeness at $17.7$ mag in $W1$ band and $17.5$ mag in $W2$ band. 
In \citep{2012ApJ...753...30S}, it is shown that a simple mid-infrared color criterion $W1-W2\geq 0.8$ identifies both un-obscured and obscured AGNs. Ref. \cite{Secrest_2015} has also selected $\approx 1.4$ million AGNs/Quasars from WISE data using two-color selection criterion. We adopted the same procedure as mentioned in \citep{Secrest:2020CPQ} to select the quasars from the catWISE2020 catalogue and used the same criterion for cuts on the data. {Various cuts \pt{to} the data are applied in order to select the best candidates that are supposed to be free from any known systematics and bias. We use {\tt HEALpix} visualization and project data with {\tt NSIDE} 64 for our analysis.}
{The final source count (left) and mask (right) maps are shown in Figure \ref{fig:Data_Mask}.}

\section{Methodology \label{sec:MultipoleRecovery}}
{We are interested in studying the dipole signal in three observables (a) number counts $N$, (b) mean spectral index $\bar{\alpha}$ and (c) mean flux density $\bar{B}$. {The dipole in number counts has already been studied earlier} \cite{Secrest:2020CPQ,Secrest:2022uvx}. In order to study the anisotropy in spectral index we consider the mean value of this variable in a small {angular} region. We use the \texttt{Healpy} pixelation scheme for our analysis. For a given pixel p we define the mean spectral index $\bar{\alpha}_{\mathrm{p}}$ as
\begin{equation}
    \bar{\alpha}_\mathrm{p} = \frac{1}{N_\mathrm{p}}\sum_{i=1}^{N_\mathrm{p}}\alpha_{i,\mathrm{p}} 
\end{equation}
here $N_\mathrm{p}$ is the total number of sources and $\alpha_{i,\mathrm{p}}$ denotes the spectral index of $i$th source in pixel p. The sum is performed over all the sources in the pixel. 
Similarly we define the mean flux density $\bar{B}_{\mathrm{p}}$ in  pixel p as
\begin{equation}
     \bar{B}_\mathrm{p} = \dfrac{1}{N_\mathrm{p}}\sum_{i=1}^{N_\mathrm{p}}B_{i,\mathrm{p}}
\end{equation}
  } 
with $B_{i,\mathrm{p}}$ denoting the flux density of $i$th source in pixel p.
In order to extract the dipole signal we use two different procedures. 
The $\chi^2$ method is described below while the details of the \texttt{Healpy} \citep{2005ApJ...622..759G,Zonca2019} method  can be found in Appendix \ref{sec:partialMethod}.

\begin{figure}
    \centering
    \includegraphics{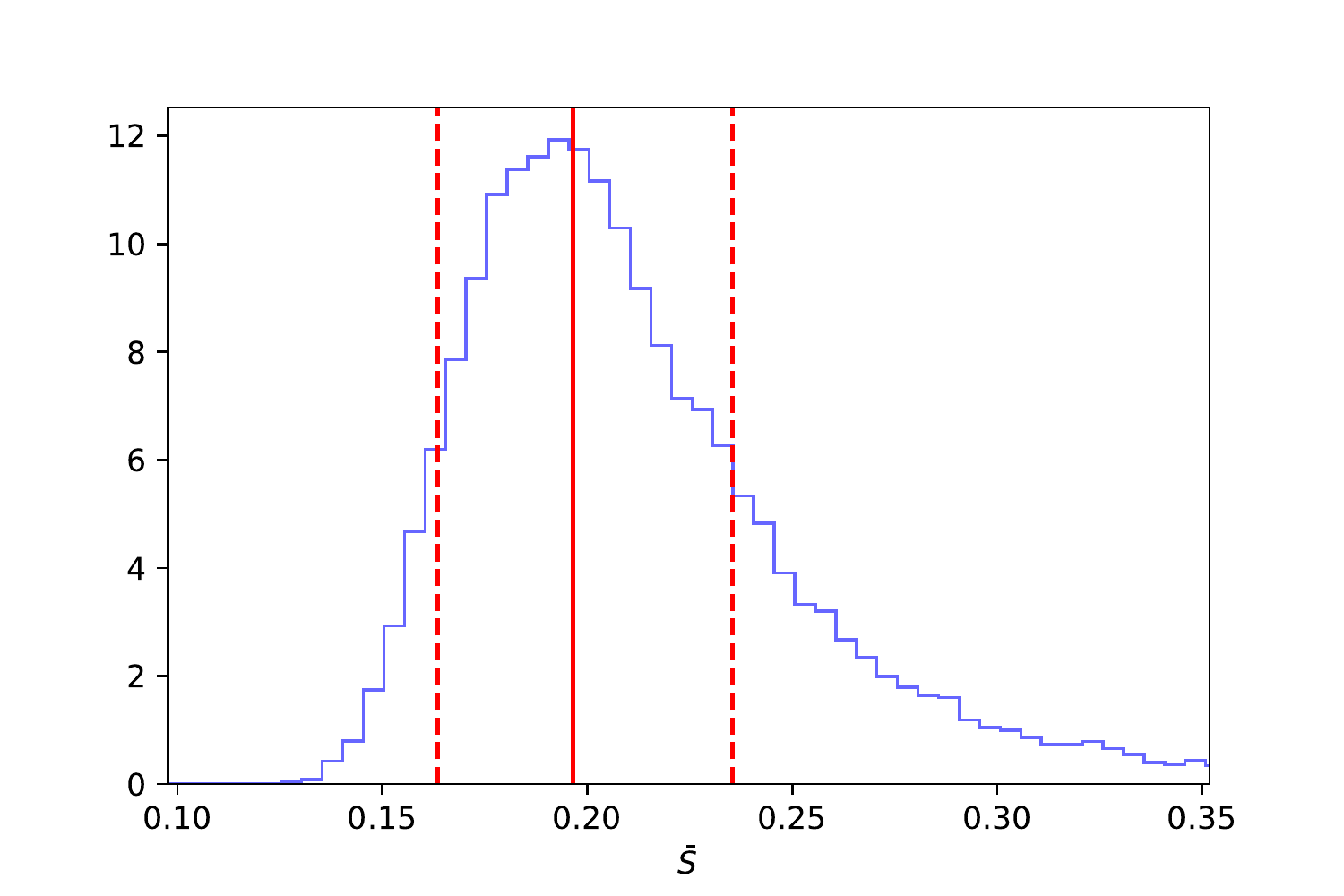}
    \caption{The mean flux density distribution for 50 sources, derived by sub-sampling the full catWISE2020 catalogue. The solid red line shows the median, and the dashed red lines are the one-$\sigma$ intervals around.}
    \label{fig:flux_PDF}
\end{figure}

\subsection{Multipole Expansion\label{sec:MultiPolExp}}
Let $I(\theta,\phi)$ denote a generic quantity of interest ($N$, $\bar{\alpha}$ or $\bar{B}$) along the direction $(\theta,\phi)$. Assuming that multipoles  $\ell\ge 3$ can be neglected, we write
\begin{equation}
	I(\theta,\phi) = \mathcal{M}_0^{\mathrm{I}}+ {\mathcal{D}}_x^{\mathrm{I}}x+{\mathcal{D}}_y^{\mathrm{I}}y+{\mathcal{D}}_z^{\mathrm{I}}z + {\mathcal{Q}}_{xy}^{\mathrm{I}}xy+ {\mathcal{Q}}_{xz}^{\mathrm{I}}xz + {\mathcal{Q}}_{yz}^{\mathrm{I}}yz + {\mathcal{Q}}_{z^2}^{\mathrm{I}}(2z^2-x^2-y^2) + {\mathcal{Q}}_{x^2-y^2}^{\mathrm{I}}(x^2-y^2)
	\label{eq:Mono_Dip_Quad_Conti}
\end{equation}
{Above} equation is {basically} the spherical harmonic decomposition in the Cartesian basis. Eq. \eqref{eq:Mono_Dip_Quad_Conti} contains 9 coefficients--one for monopole ($\mathcal{M}_0$), three for dipole $(\mathcal{D}_x, \mathcal{D}_y, \mathcal{D}_z)$, and five for quadrupole $(\mathcal{Q}_{xy}, \mathcal{Q}_{xz}, \mathcal{Q}_{yz}, \mathcal{Q}_{z^2}, \mathcal{Q}_{x^2-y^2})$, to be solved for. The superscript $I$ denotes the observable being considered. 

The observed dipole for any of the observables $\mathbf{D}$ is related to the dipole components $\bm{\mathcal{D}}_i$ present in Eq. \eqref{eq:Mono_Dip_Quad_Conti} as  
\begin{equation}
    |\mathbf{D}|=  {\sqrt{\mathcal{D}^2_x +\mathcal{D}^2_y  + \mathcal{D}^2_z}\over \mathcal{M}_0} \,.
\end{equation}

\begin{figure}
	\centering
	\includegraphics[scale=0.35]{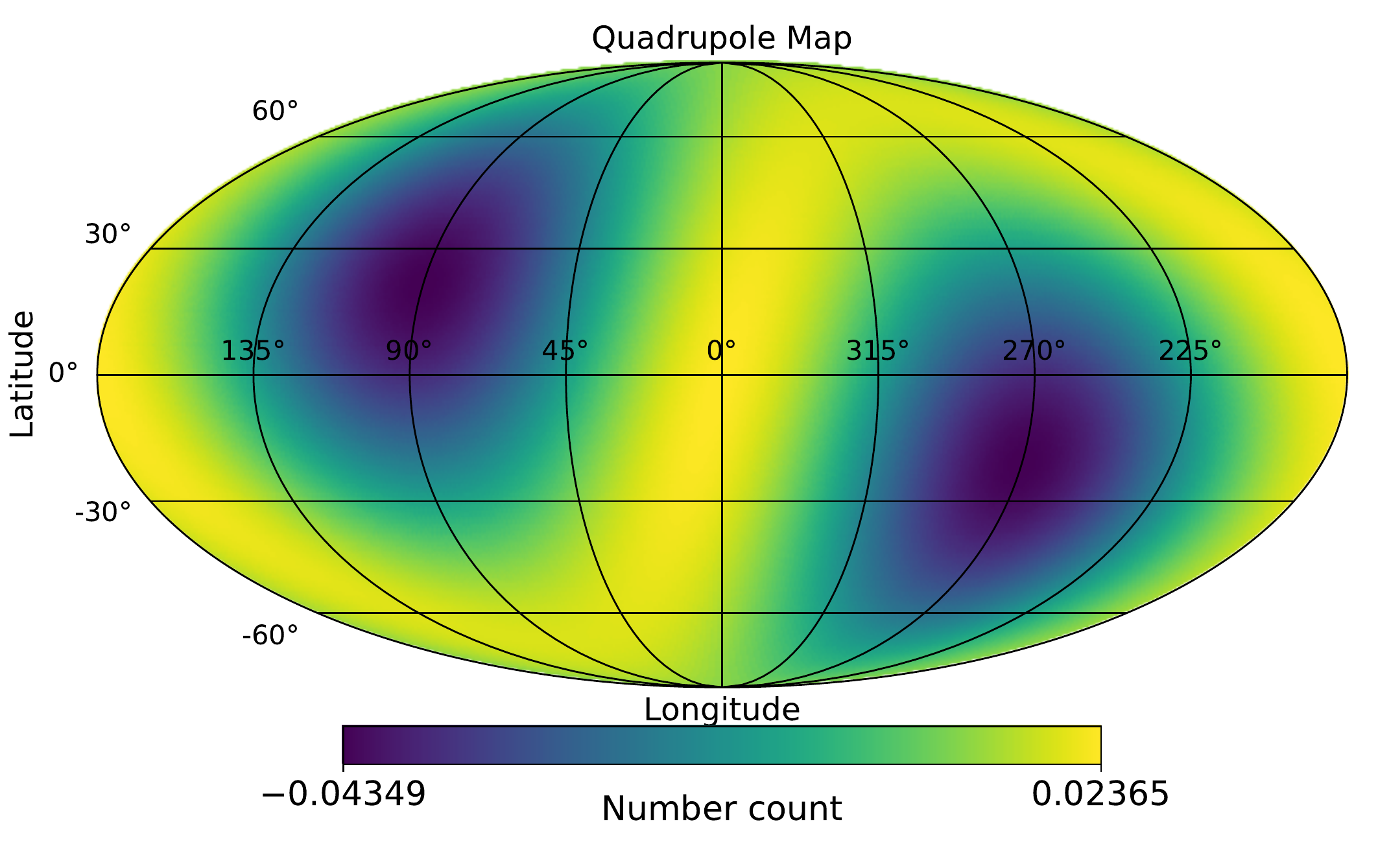}	\includegraphics[scale=0.35]{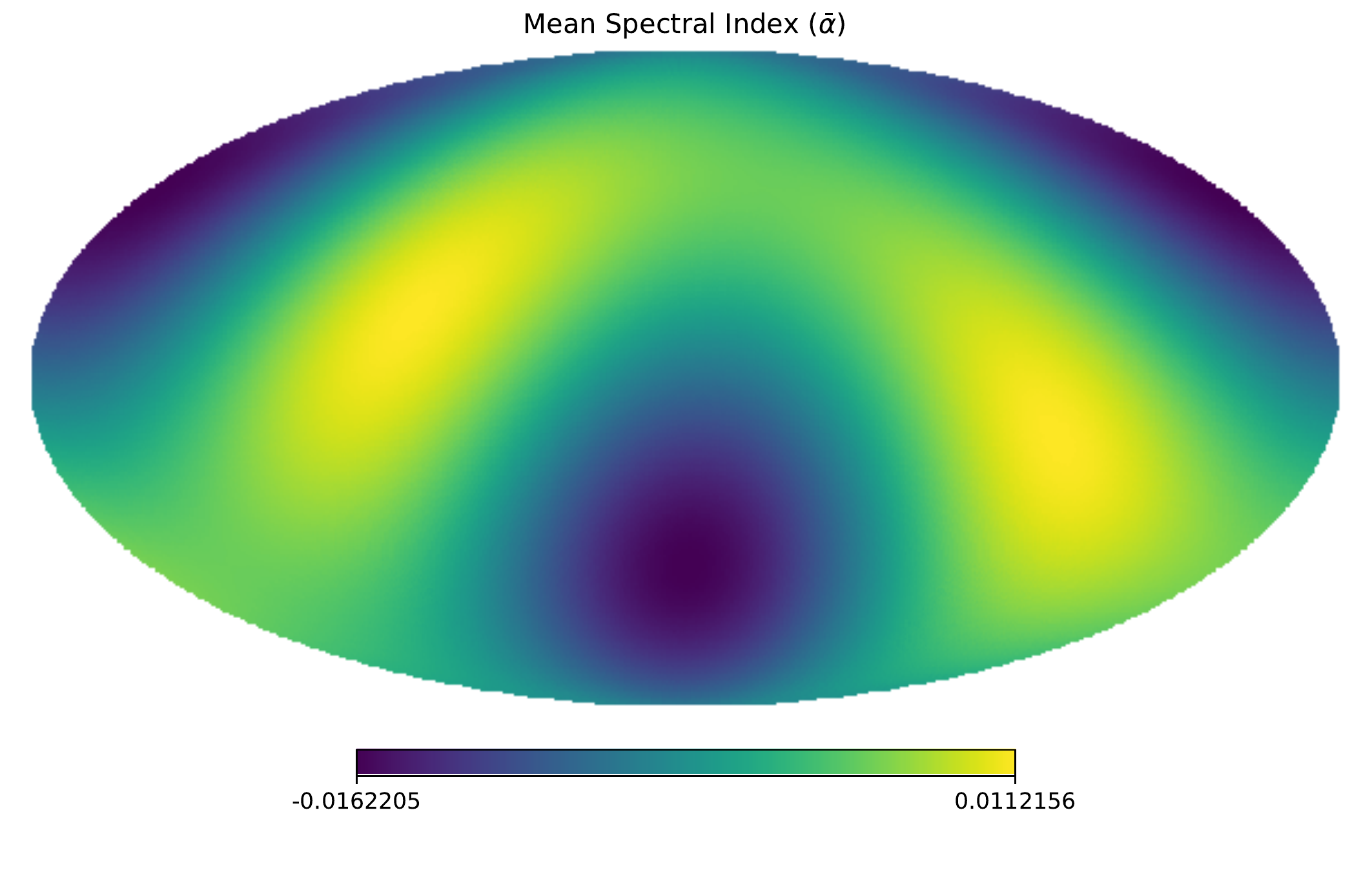}
	\includegraphics[scale=0.35]{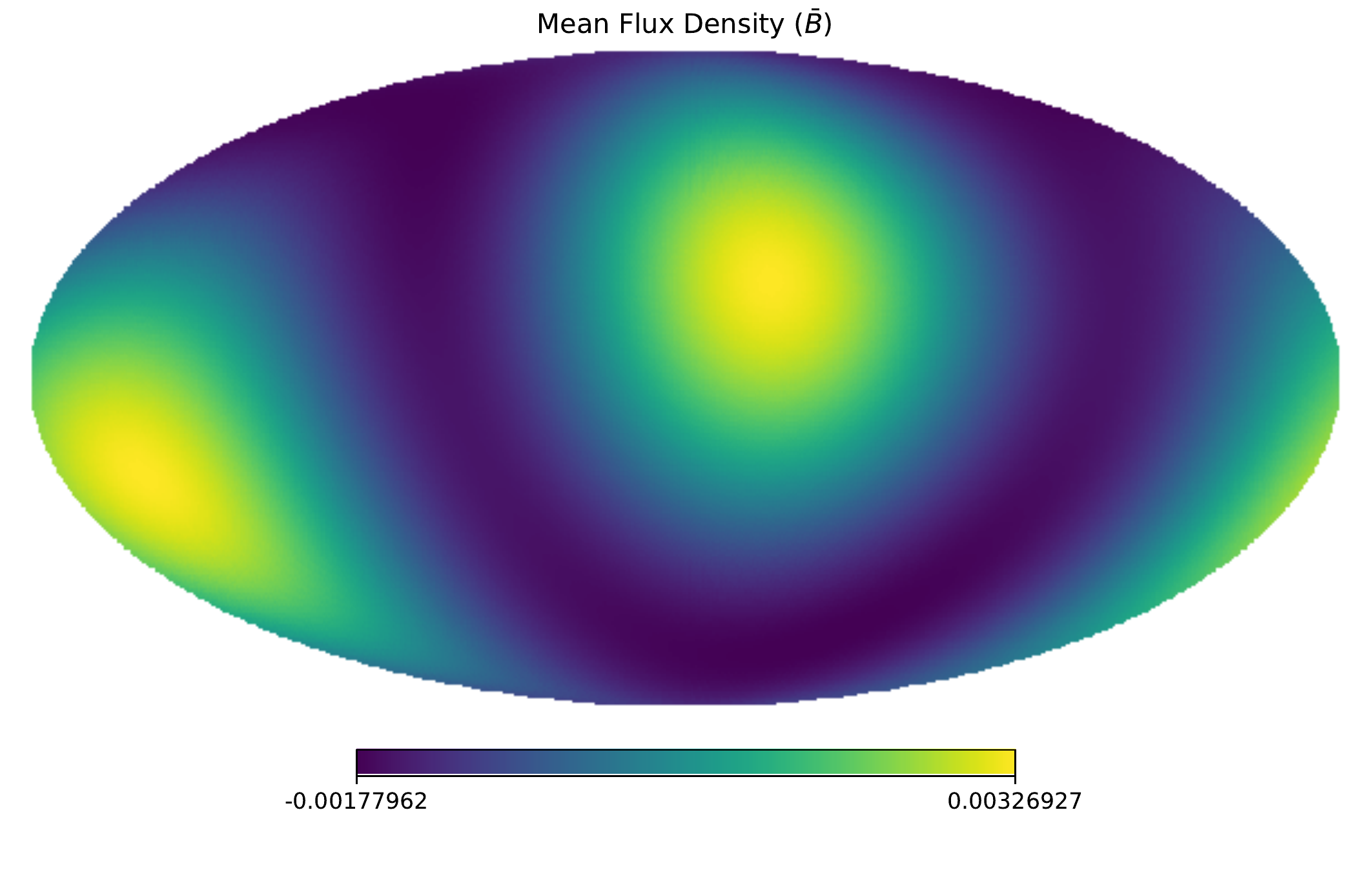}
	\caption{The extracted quadrupole for  number counts $N$, mean spectral index $\bar{\alpha}$ and mean flux density $\bar{B}$ (anticlockwise from top left)
	in the Galactic coordinate system. It can be seen from the figure, the number counts quadrupole (top row left) broadly aligns with the ecliptic poles, indicating systematic bias.}
	\label{fig:fig_quad_number}
\end{figure}

\subsection{The $\chi^2$ Statistic\label{sec:ChiSqStatisic}}
In order to determine the coefficients of Eq. \eqref{eq:Mono_Dip_Quad_Conti}, we first divide the whole sky into equal-area pixels using python package {\tt Healpy} and  determine the quantity of interest $I_\mathrm{p}$ for the pixel p. Then we determine the coefficients in Eq. \eqref{eq:Mono_Dip_Quad_Conti} using the $\chi^2$ minimization, 
\begin{equation}
	\chi^{2}=\sum_{\mathrm{p}=1}^{N_\mathrm{t}}\Bigg[\frac{I_{\mathrm{p}} - 
I(\theta,\phi)}{\sigma_{\mathrm{p}}^\mathrm{I}}\Bigg]^2\label{eq:chisq}
\end{equation}
where $\sigma_{\mathrm{p}}^\mathrm{I}$ denotes the error in the observable $I_{\mathrm{p}}$ in a given pixel p, $N_\mathrm{t}$ is the number of unmasked pixels,  and $I(\theta,\phi)$ is given in Eq. \eqref{eq:Mono_Dip_Quad_Conti}. 

{Although, the source  number count follows the Poisson distribution, on account of a large number of sources in pixels, it approaches the Gaussian distribution. Thus for the number count map, for a pixel with $N_\mathrm{p}$ sources, we consider  $\sigma_\mathrm{p}^N = \sqrt{N_\mathrm{p}}$. For observables $\bar{\alpha}$ and $\bar{B}$, the distribution is non-trivial and asymmetric. So to determine $\sigma_\mathrm{p}$ for these observables,  we resort to sub-sampling. %To determine $\sigma_\mathrm{p}$
For a pixel with $n$ sources, we draw sub-samples with a number count $n$ from the full catWISE2020 catalogue. Next, we determine the  $\bar{\alpha}$ or $\bar{B}$ of these sub-samples. The observed distribution for mean flux density is shown in Figure \ref{fig:flux_PDF}. We notice the skewness in distribution, and draw a $1\sigma$ confidence interval around the median and determine asymmetric error bars, i.e.,  $\sigma_\mathrm{p}^{(\pm)}$. Depending on whether the model's {predicted value} is larger or smaller than the observed value, we {choose $\sigma_{\mathrm{p}}^\mathrm{I}$ equal to } $\sigma_\mathrm{p}^{(+)}$ or $\sigma_\mathrm{p}^{(-)}$ and perform $\chi^2$ minimization.}

\section{Results and Discussion\label{sec:Results}}
In this section, we give the results of multipole extraction for all three observables using the $\chi^2$ method. We apply our second method to number only, as it involves symmetric error bars (see \S\ref{sec:ChiSqStatisic}).

\subsection{Source Number Counts $N$}
Using $\chi^2$ method, we find the dipole direction $(l,b)=(238.5^\circ\pm 7.8^\circ,29.6^\circ\pm5.8)$. This direction is quite close to the CMB dipole. %, the difference in angle being $\approx 26^\circ$. 
On the other hand, the magnitude of the dipole is found to be $|\mathbf{D}|=0.017\pm 0.002$. This is much larger in comparison to the expected kinematic dipole and is consistent with the value found in \cite{Secrest:2022uvx} that was obtained by a different data analysis procedure. The extracted quadrupole in number counts is shown in Figure \ref{fig:fig_quad_number} (top row left). As we can see from the figure, it broadly aligns with the ecliptic poles, indicating systematic bias as already noted in \cite{Secrest:2022uvx}. In our procedure, we do not need to model this bias since the quadrupole is directly extracted from {the} data along with the dipole.

\begin{figure}
	\centering
	\includegraphics[height=5cm]{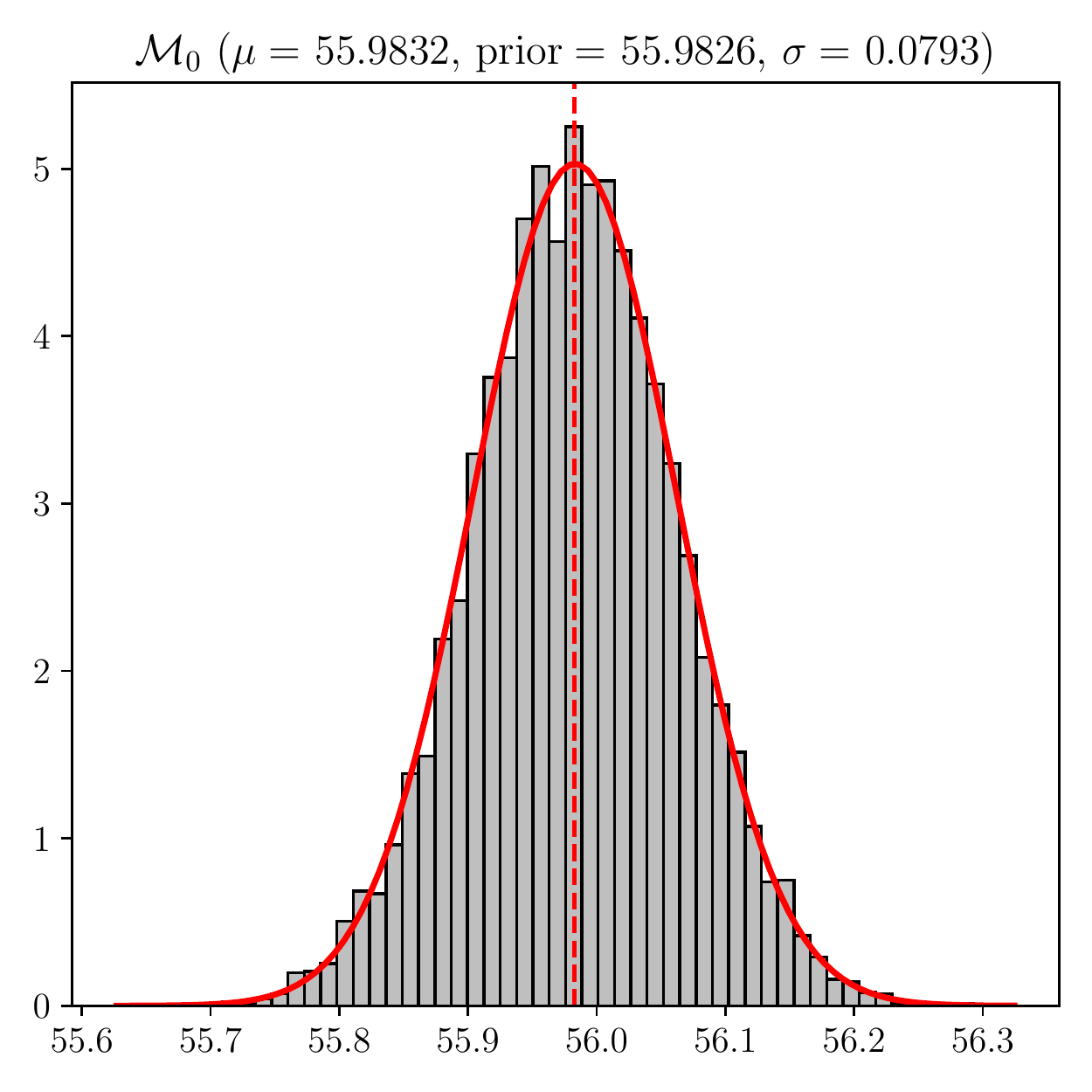}
	\includegraphics[height=5cm]{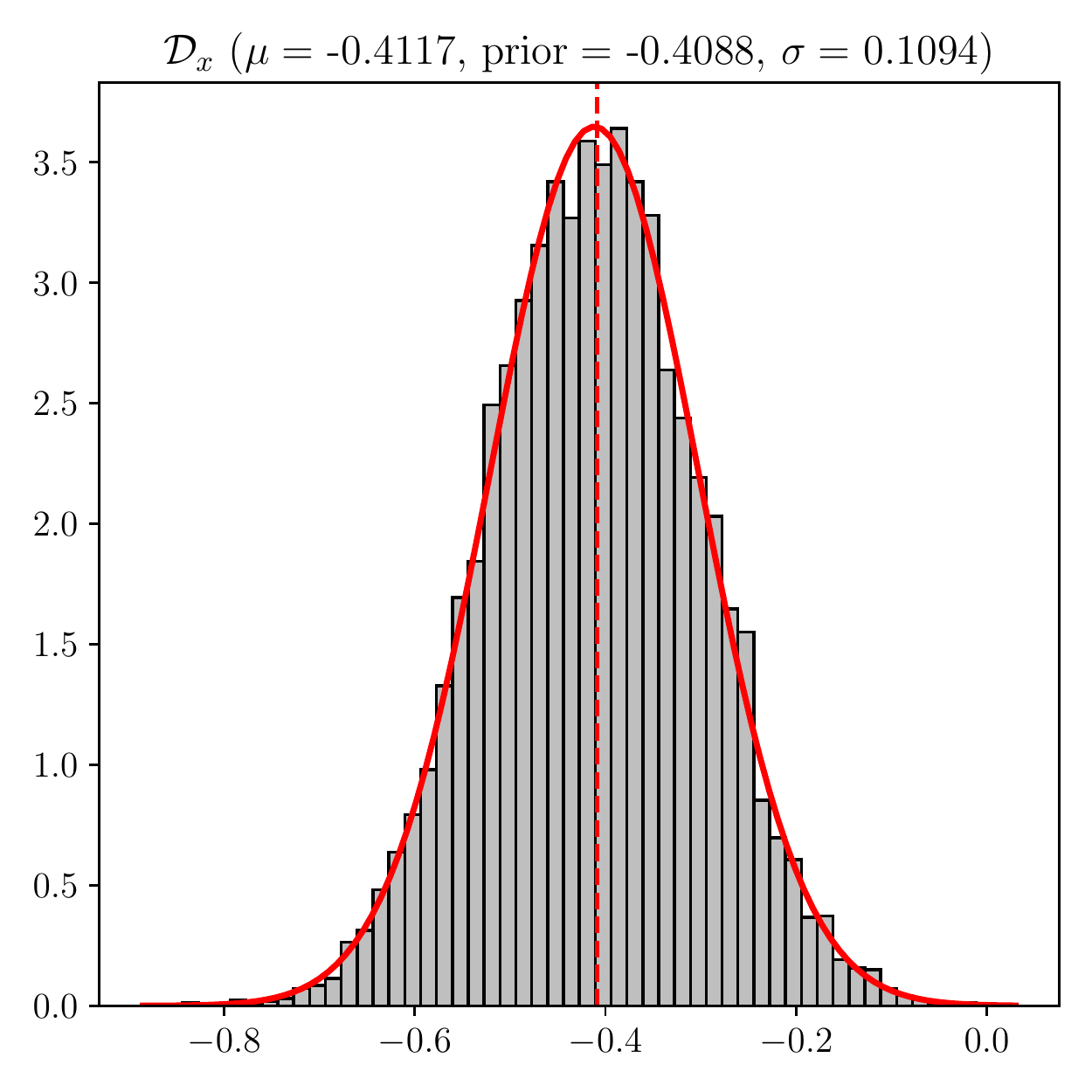}
	\includegraphics[height=5cm]{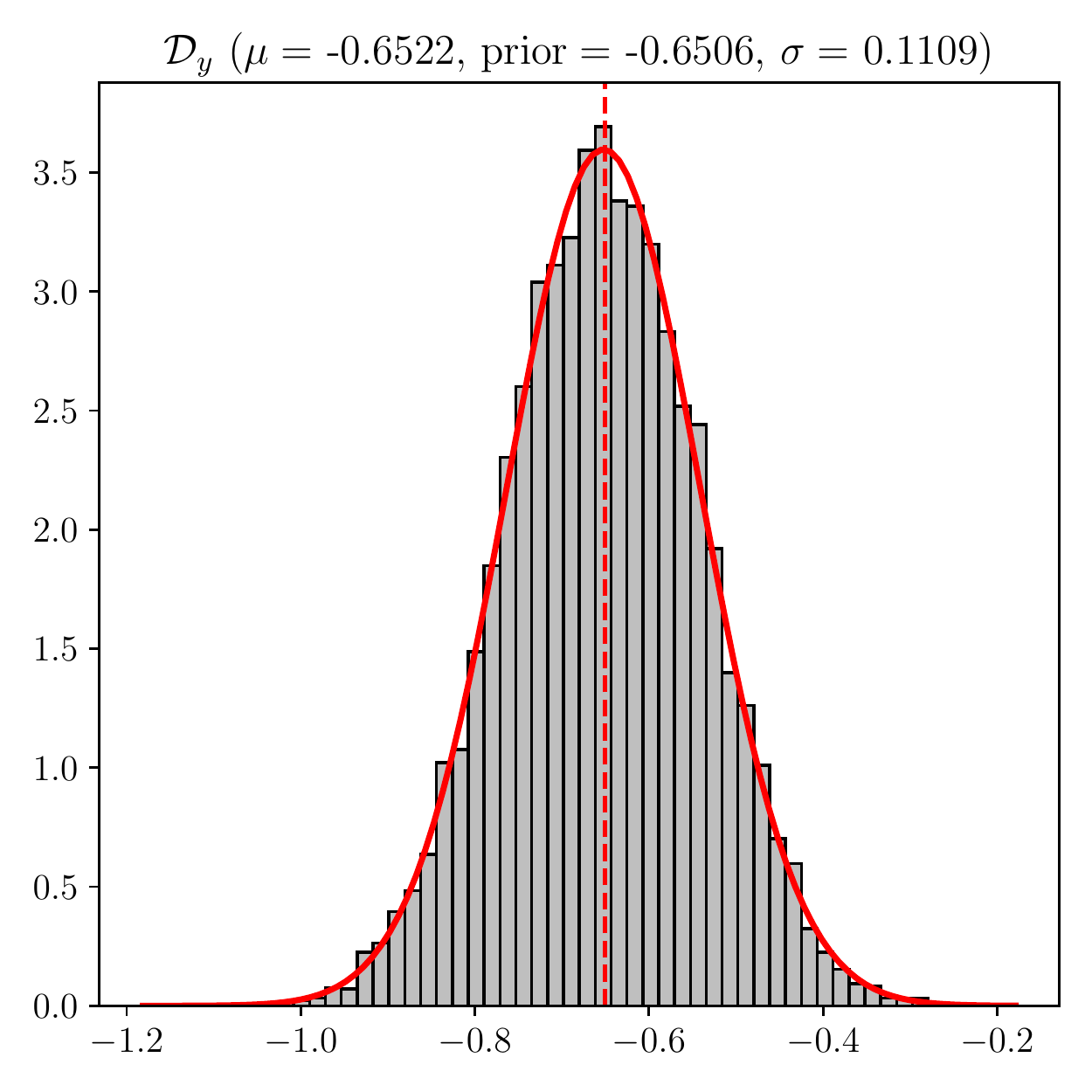}\\
	\includegraphics[height=5cm]{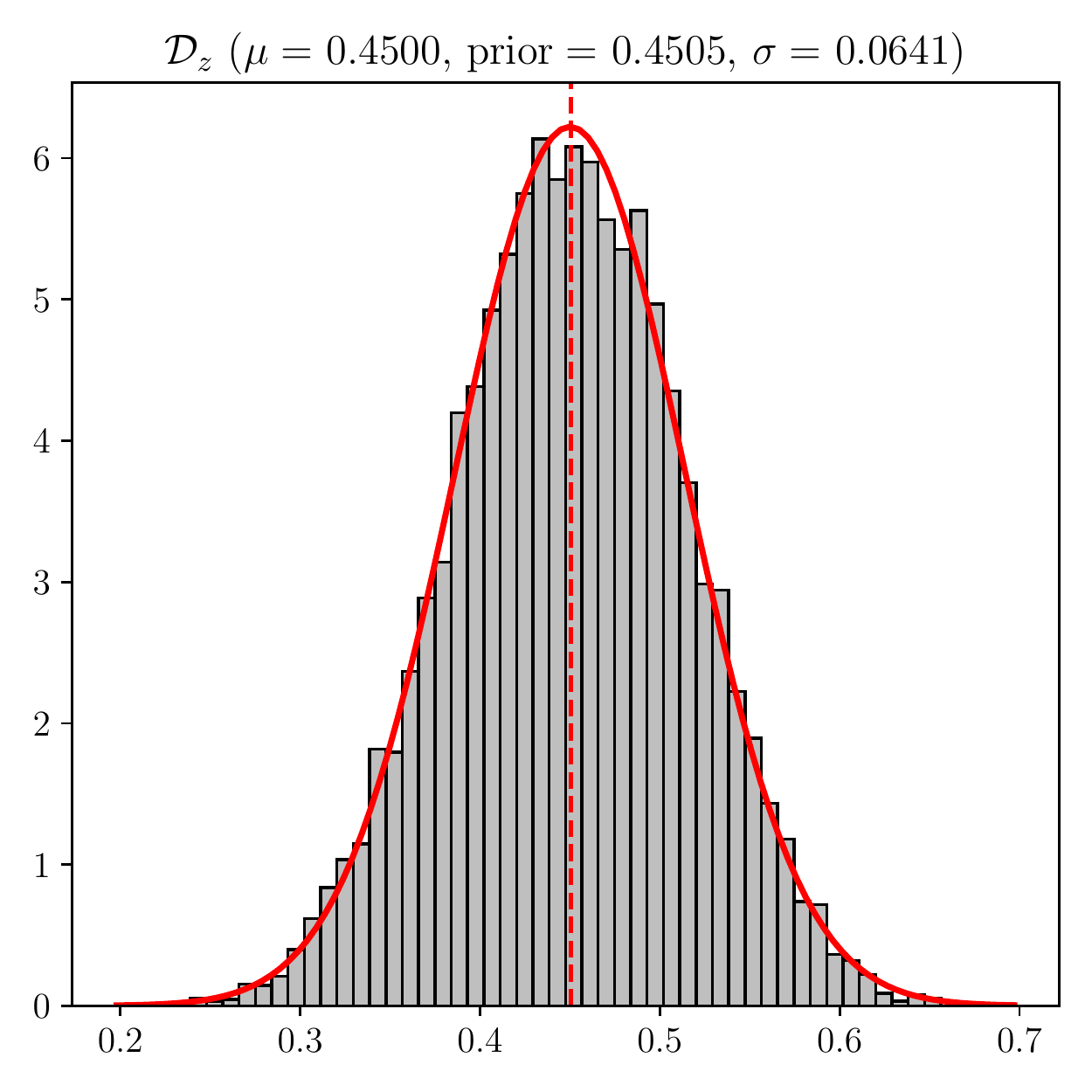}
	\includegraphics[height=5cm]{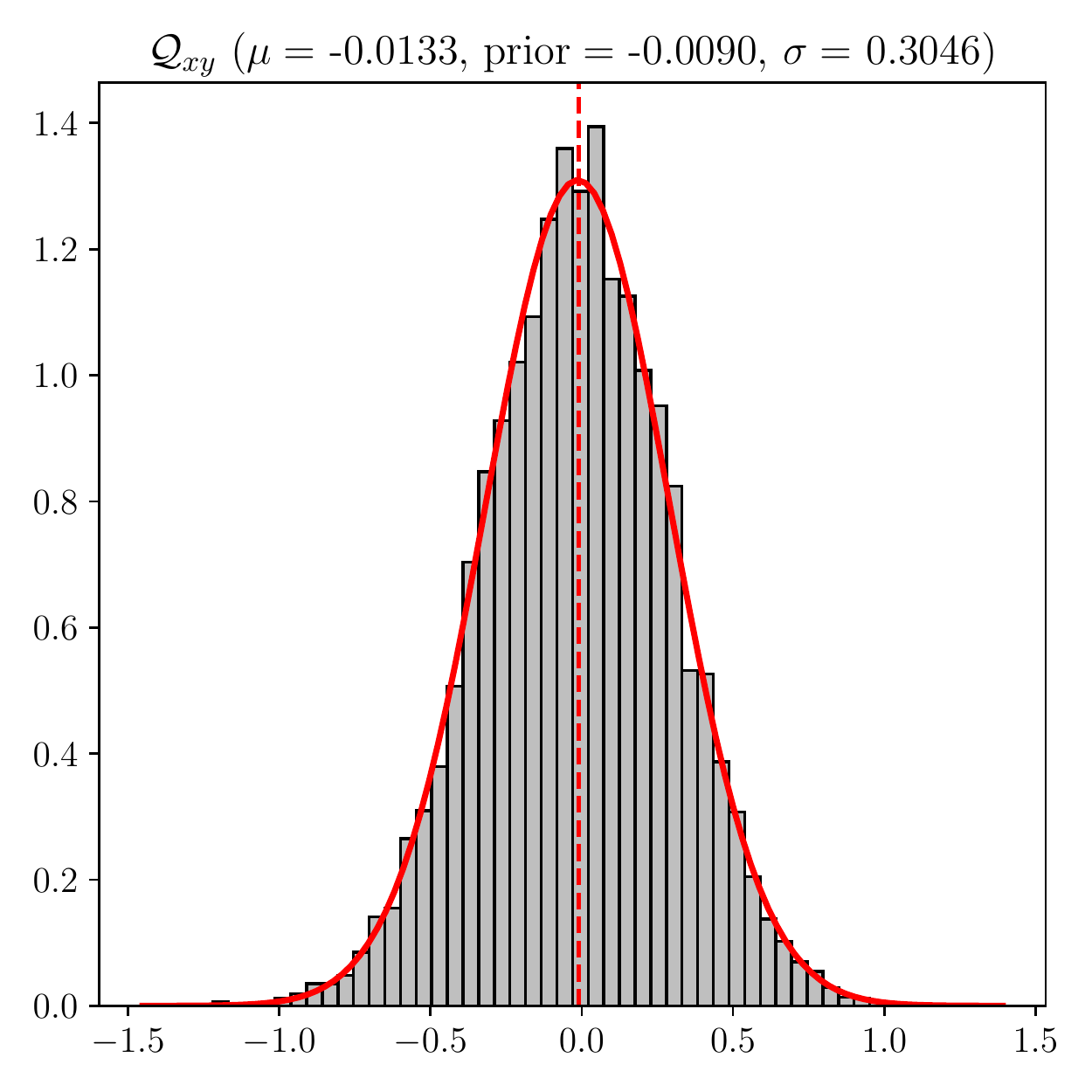}
	\includegraphics[height=5cm]{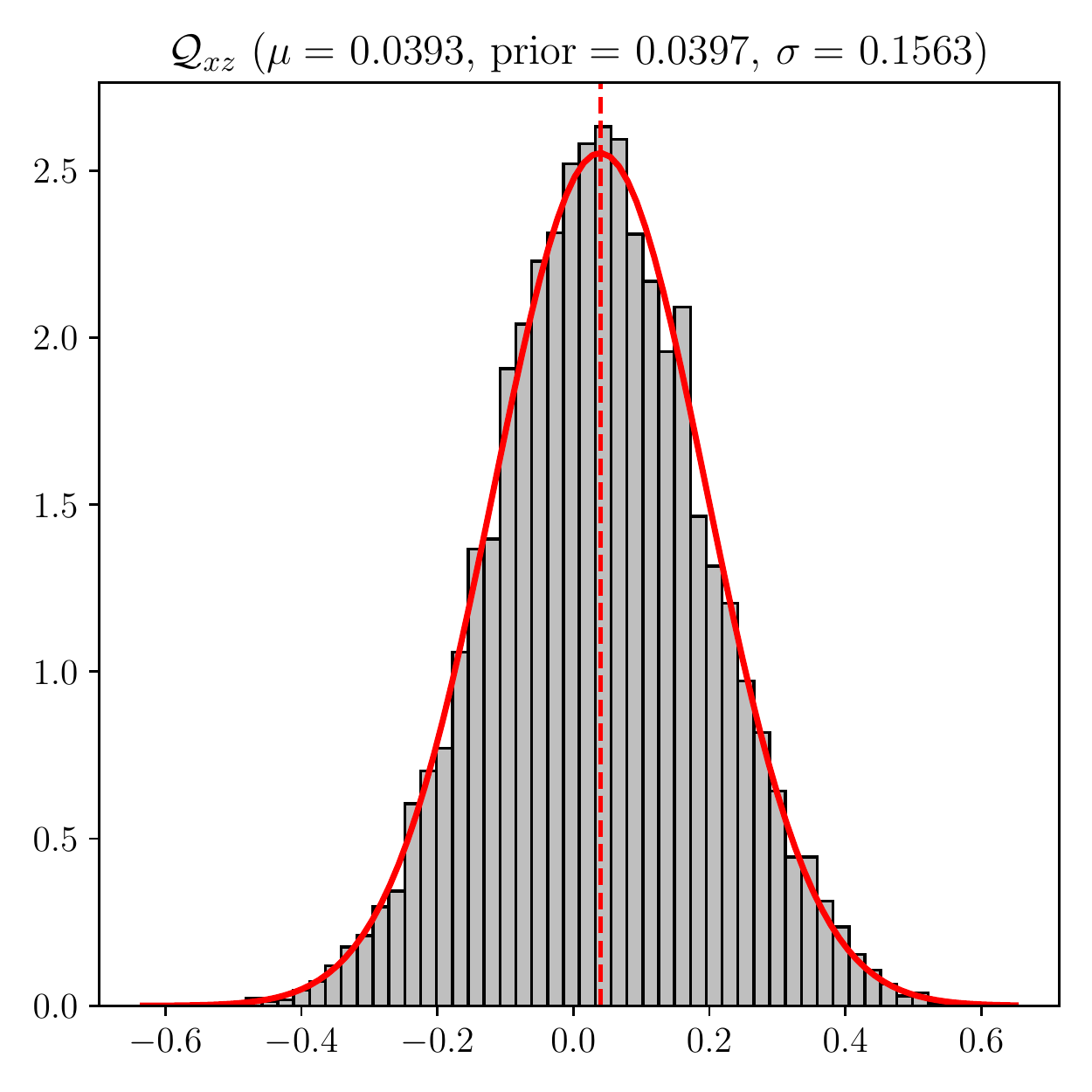}\\
	\includegraphics[height=5cm]{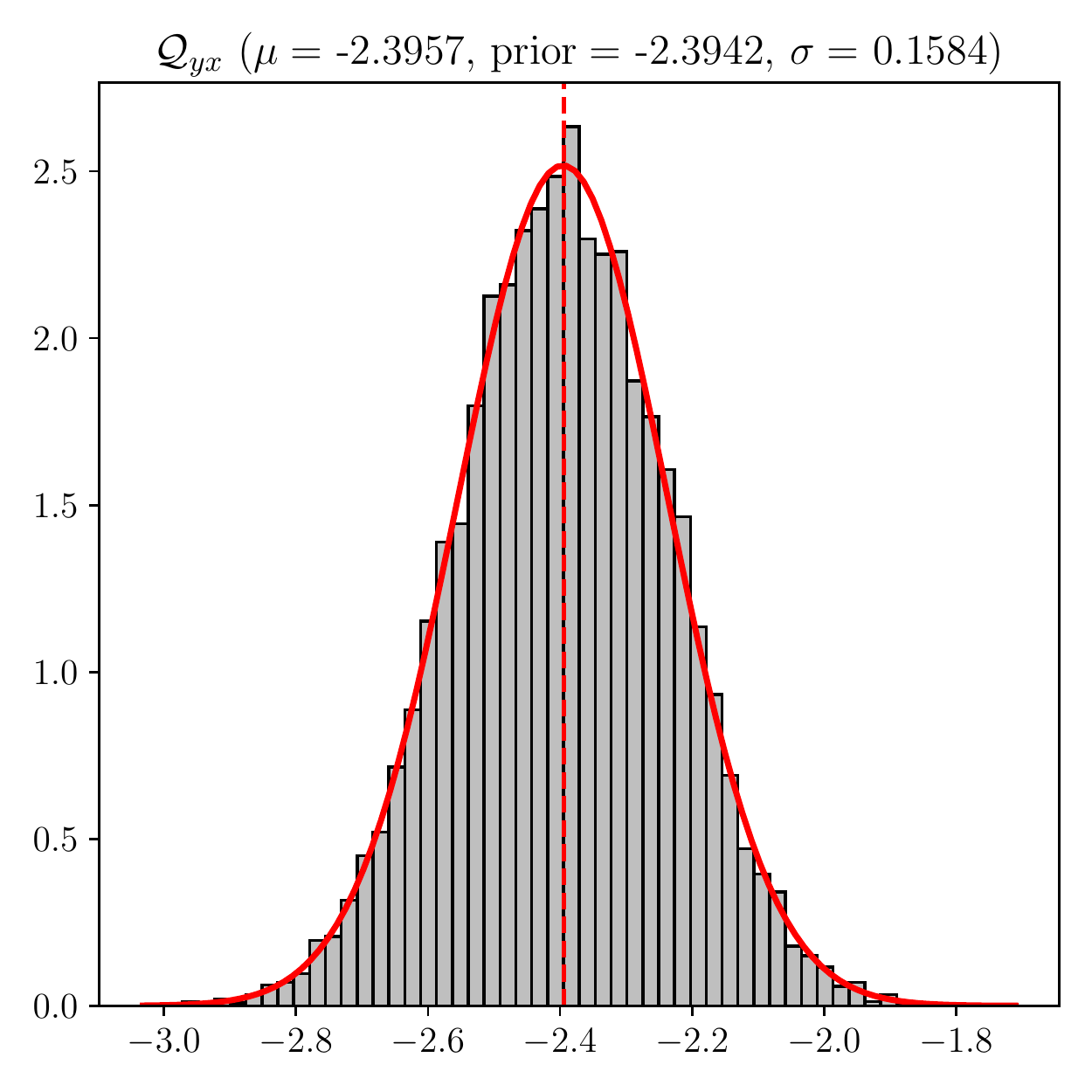}
	\includegraphics[height=5cm]{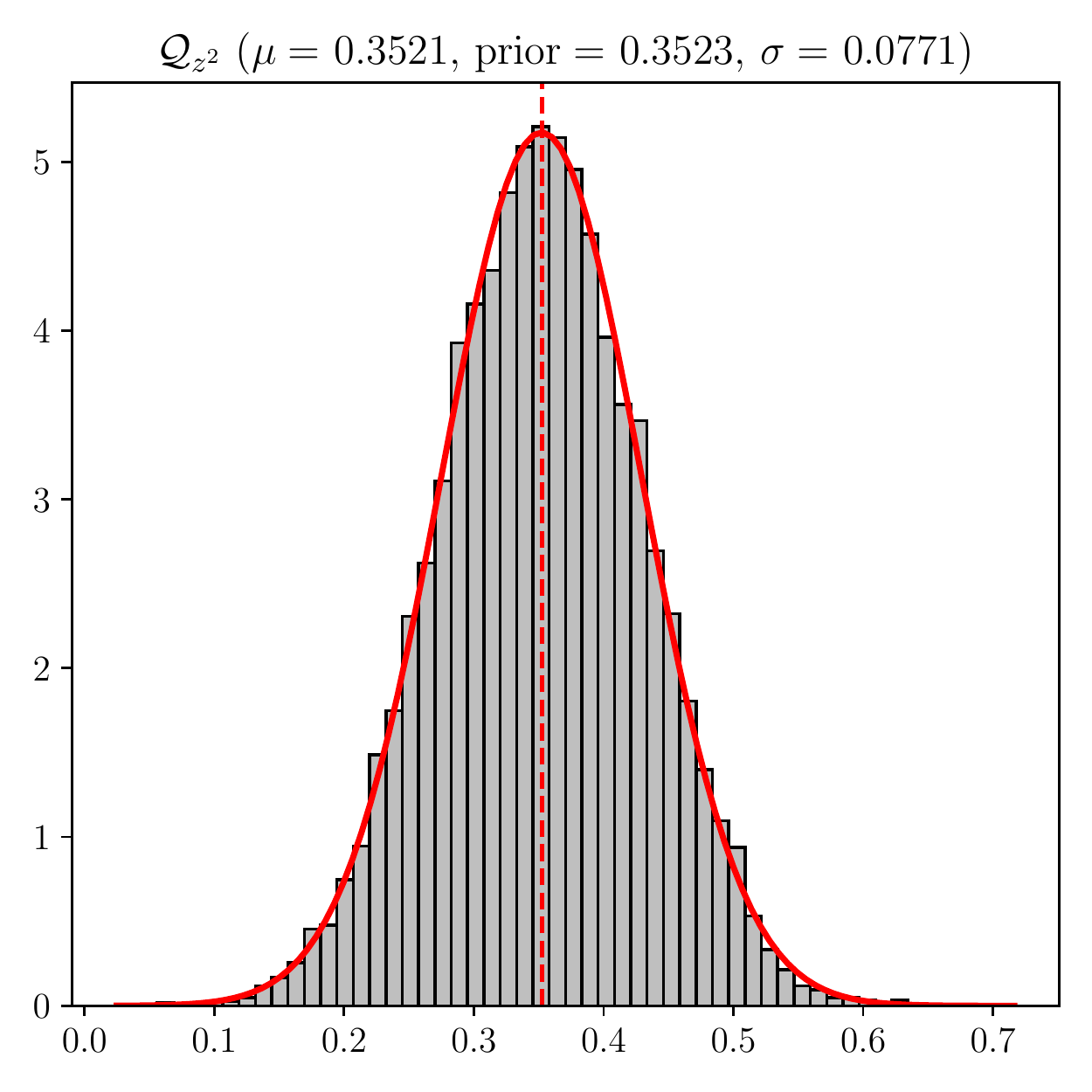}
	\includegraphics[height=5cm]{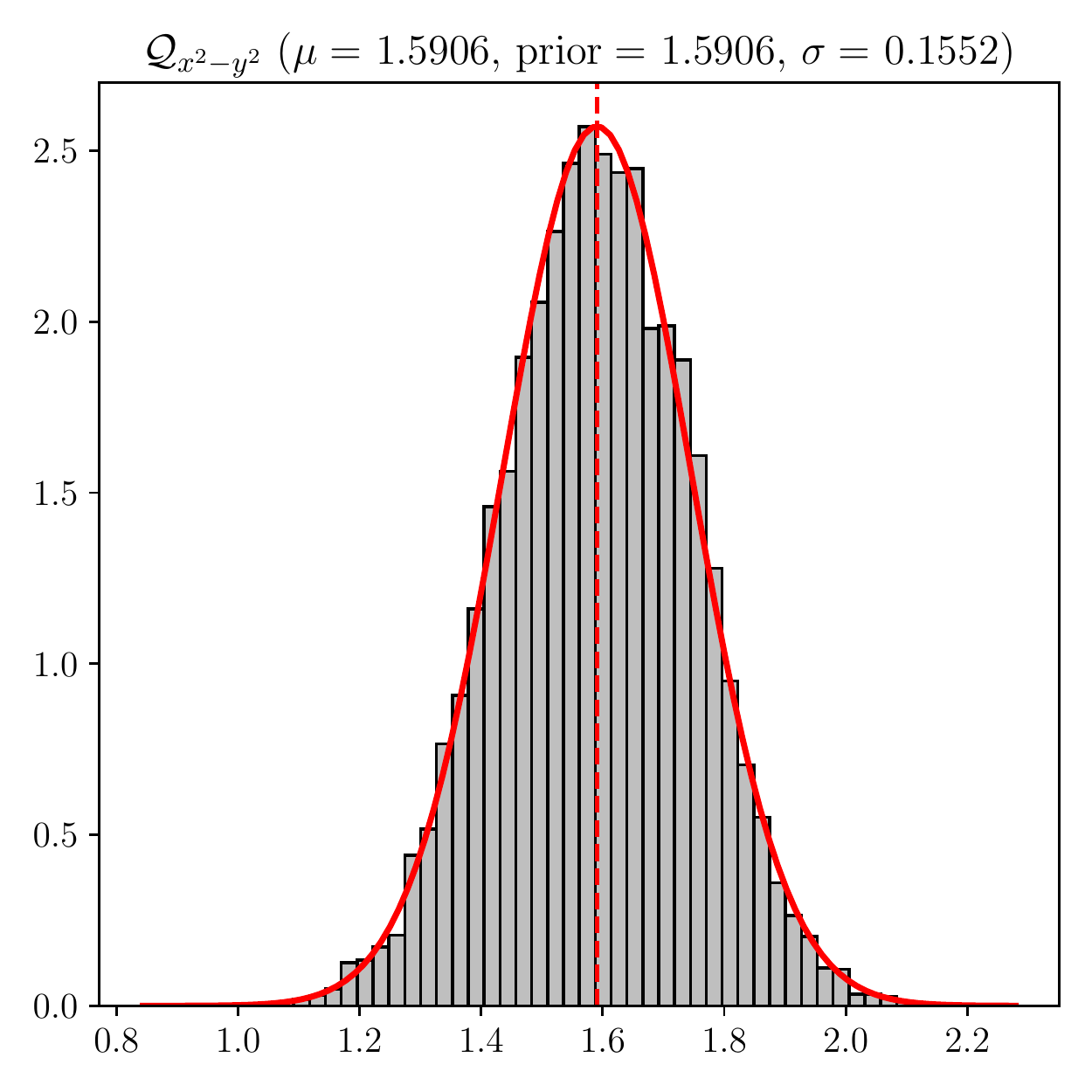}
\caption{Distribution histogram of monopole, dipole and quadrupole components (Eq. \ref{eq:Mono_Dip_Quad_Conti}) for source number counts $N$ obtained using \texttt{Healpy} method. In the plot, thick vertical dotted lines are priors and thick red curve is the Gaussian fit to the histogram. The corresponding mean ($\mu$) standard deviation ($\sigma$) are also shown in the corresponding titles.}
	\label{fig:GaussFitNumCounts}
\end{figure}

{We now make a comparison of $\chi^2$ results with those obtained using \texttt{Healpy}.  The details of the multipole extraction and error estimation can be found in Appendix \ref{sec:partialMethod}. The results are shown in Figure \ref{fig:GaussFitNumCounts}. These are the histograms of various multipole components obtained using 10,000 simulations. The dipole magnitude and directions are found to be $|\mathbf{D}|=0.016\pm 0.002$ and $(l,b)=(237.9^\circ\pm 8.4^\circ, 30.4^\circ\pm 5.0^\circ)$ respectively. Thus we find the results of \texttt{Healpy} method are quite close to those obtained using $\chi^2$ method. We have also shown the dependence of extracted monopole $\mathcal{M}_0$ on the order of expansion (Eq. \ref{eq:Mono_Dip_Quad_Conti}) in Table \ref{tab:MonComp}, using \texttt{Healpy} method.} 

It is important to ascertain that the values of extracted multipole components don't depend upon the order to which we make the expansion in Eq. \eqref{eq:Mono_Dip_Quad_Conti}. To this end, we find that $\mathcal{M}_0$ shows no change after $\ell=2$. We ascribe this to the fact that beyond $\ell=2$, the multipole components are negligible.

\begin{table}
	\centering %
	\begin{tabular}{ccccc}
		\hline
		\hline
		\noalign{\vskip 0.1cm}   & $\ell\le0$  & $\ell\le1$ & $\ell\le2$  & $\ell\le3$ \\
		\noalign{\vskip 0.1cm}
$\mathcal{M}_0$  & $56.29$  & $56.30$  & $55.98$  & $55.98$   \\
	 \hline
		\hline 
	\end{tabular}
	\caption{ Comparison of the monopole values $\mathcal{M}_0$ in number counts by considering the expansion for various $\ell$ values. The monopole value shows no change after $\ell=2$.}
	\label{tab:MonComp}
\end{table}
%Using this we can calculate the speed wrt to the Cosmic rest frame. %The speed is found to be $480\pm60$ km/s.

\subsection{{Mean} Spectral Index $\bar{\alpha}$}
We find a significant signal of dipolar anisotropy in this observable. The dipole amplitude is found to be $0.0066\pm 0.0011$
and the direction $l=171^\circ \pm 6^\circ$ and $b=7^\circ\pm 6^\circ$. %in galactic coordinates. 
%As discussed at the end of \S\ref{sec:MultiPolExp}, we use non-symmetric error bar in the $\chi^2$ analysis and subsequent multipole extraction for this observable.
As discussed in \S\ref{sec:Introduction}, this observable does not get any contribution due to kinematic effects. Thus{,} the presence of anisotropy in this parameter indicates either a bias or a possible departure from the $\Lambda$CDM. 
We also find a very strong quadrupole in the data, roughly correlated with the ecliptic poles, as shown in Figure \ref{fig:fig_quad_number} (top row right). This may arise due to observational bias which is also present in number counts.
{The dipole direction lies close to the galactic plane and points roughly opposite to the galactic center,} indicating a possible contamination from our galaxy. In order to study galactic and other sources of contamination in data, we 
{study} the effect of (a) flux cut ($S_\mathrm{cut}$) and (b) galactic cut ($b_\mathrm{cut}$) on this observable.

\subsubsection{Effect of Flux Cut}
We study the effect of flux cut $S<S_\mathrm{cut}$ as we change $S_\mathrm{cut}$ from $\infty$ to 0.1 mJy. The results are given in Table \ref{tab:fluxAlpha}. 
\begin{table}
\centering
    \begin{tabular}{p{2.0cm}p{1.5cm}p{1cm}p{3cm}p{1.8cm}p{1.8cm}}
    \hline
    \hline
    \noalign{\vskip 0.1cm}
    $S_\mathrm{cut}$ (mJy) & Sources & $f_\mathrm{sky}$ & $|\mathbf{D}|$ & $l$ (deg.) & $b$ (deg.)\\
    \noalign{\vskip 0.1cm}
    \hline
    \noalign{\vskip 0.1cm}
$\infty$ & $1307511$ & $0.4724 $ & $0.0066 \pm 0.00113$ & $171 \pm  6$ & $ 7 \pm  6$ \\
$  10.0$ & $1307023$ & $0.4724 $ & $0.0067 \pm 0.00099$ & $172 \pm  5$ & $ 9 \pm  4$ \\
$   0.7$ & $1263875$ & $0.4722 $ & $0.0067 \pm 0.00148$ & $167 \pm  9$ & $ 8 \pm  5$ \\
$   0.3$ & $1120488$ & $0.4721 $ & $0.0065 \pm 0.00090$ & $168 \pm 12$ & $10 \pm  6$ \\
$   0.2$ & $ 953664$ & $0.4718 $ & $0.0064 \pm 0.00071$ & $172 \pm 11$ & $10 \pm  4$ \\
$   0.1$ & $ 302607$ & $0.4647 $ & $0.0070 \pm 0.00372$ & $217 \pm 43$ & $22 \pm 20$ \\
    \noalign{\vskip 0.1cm}
    \hline
    \hline
    \end{tabular}
\caption{The dipole in mean spectral index $\bar{\alpha}$ as a function of flux cut $S<S_\mathrm{cut}$. It can be seen that the dipole amplitude as well as direction is almost consistent till the flux cut of 0.2 mJy. The dipole direction lies close to the galactic plane {and points opposite to the galactic center.} Beyond $S_\mathrm{cut}=0.2$, there is a drastic change in the number of sources, leading to drastic changes in the magnitude, direction and significance of the dipole.}
\label{tab:fluxAlpha}
\end{table}
\begin{table}
\centering
    \begin{tabular}{p{2.0cm}p{1.5cm}p{1cm}p{3cm}p{1.8cm}p{1.8cm}}
    \hline
    \hline
    \noalign{\vskip 0.1cm}
    $b_\mathrm{cut}$ (deg.) & Sources & $f_\mathrm{sky}$ & $|\mathbf{D}|$ & $l$ (deg.) & $b$ (deg.)\\
    \noalign{\vskip 0.1cm}
    \hline
    \noalign{\vskip 0.1cm}
$30$ & $1307511$ & $0.4724$ &  $0.0066 \pm 0.00113 $  & $171 \pm  6$  &  $  7 \pm  6$ \\
$35$ & $1117078$ & $0.4030$ &  $0.0074 \pm 0.00147 $  & $161 \pm  9$  &  $  5 \pm  6$ \\
$40$ & $ 950186$ & $0.3426$ &  $0.0055 \pm 0.00099 $  & $172 \pm 14$  &  $  4 \pm  5$ \\
$45$ & $ 767093$ & $0.2765$ &  $0.0054 \pm 0.00135 $  & $150 \pm 14$  &  $  7 \pm 10$ \\
$50$ & $ 619814$ & $0.2235$ &  $0.0036 \pm 0.00184 $  & $147 \pm 26$  &  $ 14 \pm 15$ \\
    \hline
    \hline
    \end{tabular}
\caption{The dipole in mean spectral index $\bar{\alpha}$ as a function of galactic cut. This means we have considered those sources with $|b|>b_\mathrm{cut}$. The number of sources decrease very rapidly leading to decrease in significance as the cut is increased.}
\label{tab:bCutAlpha}
\end{table}
From the table we observe that 
\begin{enumerate}
    \item Both dipole amplitude and the direction remains almost constant till $S_\mathrm{cut}=0.2$ mJy.
    \item At 0.1 mJy cut, the error in both the the magnitude and the direction increases considerably. This can be attributed to the drastic change in the number of sources from 953,664 to 302,607 as we go from 0.2 to 0.1 mJy. However, within errors, the result remains the same as  with other less stringent flux cuts.
\end{enumerate}
\subsubsection{Effect of Galactic Cut}
The dipole direction in the mean spectral index is found to be close to the galactic plane, indicating galactic contamination. {To explore this further, we} study the effect of galactic cut on the observable. This is depicted in Table \ref{tab:bCutAlpha}. From the table we can observe the following
\begin{enumerate}
    \item For the galactic cut $30^\circ\le b_\mathrm{cut}\le45^\circ$, we find a very significant dipole signal. The significance is considerably reduced for more stringent galactic cuts $ b_\mathrm{cut}\le 50^\circ$. The direction for all the cuts in the range $30^\circ\le b_\mathrm{cut}\le40^\circ$ agrees within errors. It deviates as we impose a more stringent cut $ b_\mathrm{cut}\ge 45^\circ$, but the deviation is not very significant. 
    \item The effect is not significant for more stringent galactic cuts such as $ b_\mathrm{cut}> 50^\circ$. 
\end{enumerate}
{Since the effect persists for a wide range of flux cuts, so we cannot attribute it to a few bright sources. The fact that it points roughly opposite to the galactic center indicates contamination from the galaxy.  Hence, it is reasonable to attribute the observed dipole to galactic bias and conclude that the current data is consistent with isotropy in this variable.}

\subsection{Mean {Flux Density} $\bar{B}$\label{sec:MeanFluxDensity}}
Here also we study the effect of flux and galactic cuts on the dipole. 
\subsubsection{Effect of Flux Cut}
For the flux cuts, the results are summarised in Table \ref{tab:BBar_fluxCut}. From this table, we observe the following
\begin{enumerate}
\item The dipole significance suddenly drops after $S_\mathrm{cut}=10$ mJy. This indicates that the dipole signal can be attributed to very bright sources
\item For the cases when the dipole is significant (first two rows), the dipole lies in the galactic plane pointing towards the galactic center
\item In all other cases, the dipole points away from the plane and its significance also reduces 
\end{enumerate}

\subsubsection{Effect of Galactic Cut}
These results are given in Table \ref{tab:bCutB}. From the table, we find that
\begin{enumerate}
\item For galactic cuts $b_\mathrm{cut}\le 40$, the dipole is significant. After this, the significance reduces considerably
\item For all flux cuts, the dipole points approximately towards galactic center 
\item Except $b_\mathrm{cut}= 45^\circ$ cut, the dipole lies close to the galactic plane
\end{enumerate}
{The results clearly suggest that
 dipole can be attributed to a few bright sources. Furthermore it may get contribution due to galactic contamination. }
\begin{table}
    \centering
    \begin{tabular}{p{2.0cm}p{1.5cm}p{1cm}p{3.3cm}p{2.2cm}p{2.2cm}}
    \hline
    \hline
    \noalign{\vskip 0.1cm}
    $S_\mathrm{cut}$ (mJy) & Sources & $f_\mathrm{sky}$ & $|\mathbf{D}|$ & $l$ (deg.) & $b$ (deg.)\\
    \noalign{\vskip 0.1cm}
    \hline
    \noalign{\vskip 0.1cm}
$\infty$ & $1307511$ & $0.4724 $ & $0.01205 \pm 0.003026$ & $352 \pm 60$ & $ 10\pm  8$ \\  
$10.0$ & $1307023$ & $0.4724 $ & $0.00815 \pm 0.001723$ & $356 \pm 16$ & $ 11\pm 17$ \\
$1.0$ & $1284744$ & $0.4724 $ & $0.00289 \pm 0.001281$ & $354 \pm 78$ & $ 38\pm 27$ \\
$0.7$ & $1263875$ & $0.4722 $ & $0.00231 \pm 0.000895$ & $306 \pm 89$ & $ 65\pm 59$ \\
$0.3$ & $1120488$ & $0.4721 $ & $0.00192 \pm 0.000906$ & $355 \pm 39$ & $ 42\pm 46$ \\
$0.2$ & $ 953664$ & $0.4718 $ & $0.00062 \pm 0.000544$ & $304 \pm 93$ & $ 32\pm 39$ \\
$0.1$ & $ 302607$ & $0.4647 $ & $0.00092 \pm 0.000421$ & $ 66 \pm 53$ & $-14\pm 17$ \\
    \noalign{\vskip 0.1cm}
    \hline
    \hline
    \end{tabular}
    \caption{The dipole parameters obtained using the quadrupole $\chi^2$ method for $\bar{B}$. The values in the first column indicate that only sources with flux less than the flux cuts are considered. The second column and the third columns are the number of sources left and sky fraction after applying the flux cut.}
    \label{tab:BBar_fluxCut}
\end{table}

\begin{table}
\centering
    \begin{tabular}{p{2.0cm}p{1.5cm}p{1cm}p{3.3cm}p{2.2cm}p{2.2cm}}
    \hline
    \hline
    \noalign{\vskip 0.1cm}
    $b_\mathrm{cut}$ (deg.) & Sources & $f_\mathrm{sky}$ & $|\mathbf{D}|$ & $l$ (deg.) & $b$ (deg.)\\
    \noalign{\vskip 0.1cm}
    \hline
    \noalign{\vskip 0.1cm}
$30$ & $1307511$ & $0.4724$ &  $0.0121 \pm 0.00303 $  & $352 \pm 60$  &  $ 10 \pm  8$ \\ 
$35$ & $1117078$ & $0.4030$ &  $0.0151 \pm 0.00165 $  & $357 \pm  2$  &  $  7 \pm  2$ \\ 
$40$ & $ 950186$ & $0.3426$ &  $0.0104 \pm 0.00177 $  & $ 23 \pm  8$  &  $ 18 \pm  5$ \\ 
$45$ & $ 767093$ & $0.2765$ &  $0.0045 \pm 0.00175 $  & $351 \pm 149$  &  $ 37 \pm 22$ \\ 
$50$ & $ 619814$ & $0.2235$ &  $0.0136 \pm 0.01535 $  & $304 \pm 53$  &  $ 21 \pm 35$ \\ 
    \hline
    \hline
    \end{tabular}
\caption{The dipole in mean flux density $\bar{B}$ as a function of galactic cut. Whenever the dipole is significant, the direction lies close to the galactic plane pointing towards galactic center.}
\label{tab:bCutB}
\end{table}

\section{Conclusion and Outlook \label{sec:Conclusion}}
In this paper, we have studied the dipole in three observables --- source number counts, {mean} spectral index, and mean {flux density}. We have used two different data analysis methods ($\chi^2$ and \texttt{Healpy}) which directly extract the first three multipoles  {($\ell=0,1,2$)} and the corresponding errors from {the} data. The higher multipoles are neglected since they are found to be small. %The errors in the extracted parameters are estimated as well. 
The \texttt{Healpy} method is applied only for number counts as it involves symmetric error bars. We find results obtained using both methods almost the same. Further, these values are found to be consistent with \cite{Secrest:2020CPQ}. This provides an independent check on their results. We point out that in our case, we do not model the bias in data associated with the ecliptic pole. This bias leads to a strong quadrupole and is directly extracted from data.    

Although the observable $N$ has symmetric error bars, this is no longer true for $\bar{\alpha}$ and $\bar{B}$. Thus we have used only $\chi^2$ method for these observables. Generalization of the \texttt{Healpy} method to the non-symmetric error bars can be pursued in future.
We find a very strong signal of dipole anisotropy in the {mean} spectral index $\bar{\alpha}$. The direction in this case lies close to the galactic plane and points roughly opposite to the galactic center. Hence to evaluate the effect of galactic contamination, we study the dipolar anisotropy as a function of the galactic cut. We find that the signal remains very significant for galactic cut
$b_\mathrm{cut} > 45^\circ $ but starts to loose significance for more stringent cuts. 
We also study the effect of flux cuts $S<S_\mathrm{cut}$. 
We find that both the amplitude and direction of the dipole don't show much change as the flux cut is made more stringent. However, as expected, the errors in the dipole parameters become very large for the very stringent flux cut of $ S_\mathrm{cut} $ = 0.1 mJy. Since the dipole in this variable points roughly opposite to the galactic center, we conclude that it may be attributed to contamination from our galaxy.
 This variable also shows a strong quadrupole roughly correlated with the ecliptic poles, which may indicate the presence of observational bias in data similar to that present in number counts. 
 
The {mean flux density} $\bar{B}$ also shows a significant dipole with direction pointing towards the galactic center. 
The dipole becomes considerably reduced and also doesn't remain significant if we impose the flux cut $S_\mathrm{cut} < 1$ mJy. Additionally, the significance of the dipole reduces beyond the galactic cut $b_\mathrm{cut}\le 40^\circ$. 
Hence it is reasonable to attribute the observed dipole in this variable to a few bright sources. In conclusion,
both the mean spectral index and the mean flux density are consistent with isotropy.

\section*{Acknowledgements}
We acknowledge the use of \texttt{python} packages \texttt{scipy} \citep{2020SciPy-NMeth}, \texttt{matplotlib} \citep{Hunter:2007} and \texttt{numpy} \citep{harris2020array} for our analysis. Rahul Kothari is supported by the South African Radio Astronomy Observatory and the National Research Foundation (Grant No. 75415). PT acknowledges the support of the RFIS grant (No. 12150410322) by the National Natural Science Foundation of China (NSFC).

\appendix
\section{An alternative method for extracting masked sky multipole coefficients  \label{sec:partialMethod}} 
In this section, we give details of an alternate procedure to extract the multipole coefficients for a given observable. Here, we have applied the method only for number counts map $N$. First, we talk about extracting the coefficients and then a method for estimating the corresponding errors. {In order to better understand the \textit{modus operandi}, we have considered some special cases. We have also given an example of our procedure in Figure \ref{fig:GaussFitNumCounts} that shows the distribution of various parameters which is found to be Gaussian to a very good extend. Additionally, we have given corresponding mean values ($\mu$) and standard deviation ($\sigma$) for number counts $N$.}

\subsection{Coefficients' Extraction}
The method is based on solving a system of linear equations. In a multipole expansion, considered till $\ell$, we'd need $(\ell+1)^2$ equations for obtaining all the coefficients in Eq. \eqref{eq:Mono_Dip_Quad_Conti}. Thus in our case, we need to solve a system of 9 linear equations.  Although, in our analysis, we have terminated the expansion at $\ell=2$, yet we must emphasize that the procedure can be generalized to any order.  As the first step, we write the discretized version of Eq. \eqref{eq:Mono_Dip_Quad_Conti} {(here $r_\mathrm{p}^2=x_\mathrm{p}^2+y_\mathrm{p}^2+z_\mathrm{p}^2$)}
\begin{eqnarray}
    I_\mathrm{p} &=& \mathcal{M}_0^\mathrm{I}+ \mathcal{D}_x^\mathrm{I}\,x_\mathrm{p}+\mathcal{D}_y^\mathrm{I}\,y_\mathrm{p}+\mathcal{D}_z^\mathrm{I}\,z_\mathrm{p} + \mathcal{Q}_{xy}^\mathrm{I}\,x_\mathrm{p}y_\mathrm{p}+ \mathcal{Q}_{xz}^\mathrm{I}\,x_\mathrm{p}z_\mathrm{p}\notag \\ &+& \mathcal{Q}_{yz}^\mathrm{I}\,y_\mathrm{p}z_\mathrm{p} + \mathcal{Q}_{z^2}^\mathrm{I}\,(3z^2_\mathrm{p}-r^2_\mathrm{p}) + \mathcal{Q}_{x^2-y^2}^\mathrm{I}\,(x^2_\mathrm{p}-y^2_\mathrm{p})\label{eq:mono_dip_quad_dis}
\end{eqnarray}
In this equation, $I_\mathrm{p}$ and $(x_\mathrm{p},y_\mathrm{p},z_\mathrm{p}) $ respectively denote the observable value and cartesian coordinates of the given pixel in the \texttt{Healpy} pixelation scheme. For obtaining these coefficients, we now set up a system of linear equations. To get our first equation, we sum over all the unmasked pixels $N_\mathrm{t}$ on both sides to get
\begin{eqnarray}
    \sum_{\mathrm{p}=1}^{N_\mathrm{t}}I_\mathrm{p} &=& \mathcal{M}_0^\mathrm{I}\sum_{\mathrm{p}=1}^{N_\mathrm{t}}1 + \mathcal{D}_x^\mathrm{I}\sum_{\mathrm{p}=1}^{N_\mathrm{t}}x_\mathrm{p}+\mathcal{D}_y^\mathrm{I}\sum_{\mathrm{p}=1}^{N_\mathrm{t}}y_\mathrm{p}+\mathcal{D}_z^\mathrm{I}\sum_{\mathrm{p}=1}^{N_\mathrm{t}}z_\mathrm{p}+ \mathcal{Q}_{xy}^\mathrm{I}\sum_{\mathrm{p}=1}^{N_\mathrm{t}}x_\mathrm{p}y_\mathrm{p} + \mathcal{Q}_{xz}^\mathrm{I}\sum_{\mathrm{p}=1}^{N_\mathrm{t}}x_\mathrm{p}z_\mathrm{p} \notag\\
    &+& \mathcal{Q}_{yz}^\mathrm{I}\sum_{\mathrm{p}=1}^{N_\mathrm{t}}y_\mathrm{p}z_\mathrm{p}+\mathcal{Q}_{z^2}^\mathrm{I}\sum_{\mathrm{p}=1}^{N_\mathrm{t}}(3z^2_\mathrm{p}-r^2_\mathrm{p}) + \mathcal{Q}_{x^2-y^2}^\mathrm{I}\sum_{\mathrm{p}=1}^{N_\mathrm{t}}(x^2_\mathrm{p}-y^2_\mathrm{p})
\end{eqnarray}
To get the next equation, we multiply \eqref{eq:mono_dip_quad_dis} by $x_\mathrm{p}$ and then perform the sum. We repeat this process, by multiplying next with $y_\mathrm{p}$ and so on, i.e., all the direction dependent factors, multiplying multipole coefficients in Eq. \eqref{eq:mono_dip_quad_dis}. Finally, we get a system of 9 linear equations which can be written as
\begin{equation}
    \mathbb{M}\,\mathbb{P}=\mathbb{N}\label{eq:Fitter}
\end{equation}
here the parameter vector $\mathbb{P}$ is 
\begin{equation}
    \mathbb{P} = \big[\mathcal{M}_0^\mathrm{I}\ \mathcal{D}_x^\mathrm{I}\ \mathcal{D}_y^\mathrm{I}\  \mathcal{D}_z^\mathrm{I}\  \mathcal{Q}_{xy}^\mathrm{I}\ \mathcal{Q}_{xz}^\mathrm{I}\ \mathcal{Q}_{yz}^\mathrm{I}\ \mathcal{Q}_{z^2}^\mathrm{I}\ \mathcal{Q}_{x^2-y^2}^\mathrm{I}\big]^T
\end{equation}
and $T$ in the superscript denotes the transpose. The vector $\mathbb{N}$ on the RHS is given by
\begin{equation}
    \mathbb{N} = \Bigg[\sum_\mathrm{p}^{N_\mathrm{t}} (I_\mathrm{p}\  I_\mathrm{p}x_\mathrm{p}\ I_\mathrm{p}y_\mathrm{p}\ I_\mathrm{p}z_\mathrm{p}\ I_\mathrm{p}x_\mathrm{p}y_\mathrm{p}\ I_\mathrm{p}x_\mathrm{p}x_\mathrm{p}\ I_\mathrm{p}y_\mathrm{p}z_\mathrm{p}\ I_\mathrm{p}Z_\mathrm{p}\ I_\mathrm{p}M_\mathrm{p})\Bigg]^T
\end{equation}
where for brevity sake we have defined $Z_\mathrm{p}=3z_\mathrm{p}^2-1$, $M_\mathrm{p}=x_\mathrm{p}^2-y_\mathrm{p}^2$ and the sum outside the parenthesis is meant to be performed on all the terms inside the parenthesis over pixel p. Finally, the symmetric matrix $\mathbb{M}$ is given by the following expression
\begin{equation}
\mathbb{M}=\sum_\mathrm{p}^{N_\mathrm{t}}\begin{bmatrix}
1 & x_\mathrm{p} & y_\mathrm{p} & z_\mathrm{p} & x_\mathrm{p}y_\mathrm{p} & x_\mathrm{p}z_\mathrm{p} & y_\mathrm{p}z_\mathrm{p} & Z_\mathrm{p} & M_\mathrm{p} \\
	x_\mathrm{p} &  x_\mathrm{p}^2 & x_\mathrm{p}y_\mathrm{p} & x_\mathrm{p}z_\mathrm{p} & x_\mathrm{p}^2y_\mathrm{p} & x_\mathrm{p}^2z_\mathrm{p} & x_\mathrm{p}y_\mathrm{p}z_\mathrm{p} & x_\mathrm{p}Z_\mathrm{p} & x_\mathrm{p}M_\mathrm{p} \\
	y_\mathrm{p} &  x_\mathrm{p}y_\mathrm{p} & y_\mathrm{p}^2 & y_\mathrm{p}z_\mathrm{p} & x_\mathrm{p}y_\mathrm{p}^2 & x_\mathrm{p}y_\mathrm{p}z_\mathrm{p} & y_\mathrm{p}^2z_\mathrm{p} & y_\mathrm{p}Z_\mathrm{p} & M_\mathrm{p} \\
z_\mathrm{p} &	  x_\mathrm{p}z_\mathrm{p} & y_\mathrm{p}z_\mathrm{p} & z_\mathrm{p}^2 & x_\mathrm{p}y_\mathrm{p}z_\mathrm{p} & x_\mathrm{p}z_\mathrm{p} & y_\mathrm{p}z_\mathrm{p}^2 & z_\mathrm{p}Z_\mathrm{p} & z_\mathrm{p}M_\mathrm{p} \\
x_\mathrm{p}y_\mathrm{p} &	  x_\mathrm{p}^2y_\mathrm{p} & x_\mathrm{p}y_\mathrm{p}^2 & x_\mathrm{p}y_\mathrm{p}z_\mathrm{p} & (x_\mathrm{p}y_\mathrm{p})^2 & x_\mathrm{p}^2y_\mathrm{p}z_\mathrm{p} & x_\mathrm{p}y_\mathrm{p}^2z_\mathrm{p} & x_\mathrm{p}y_\mathrm{p}Z_\mathrm{p} & x_\mathrm{p}y_\mathrm{p}M_\mathrm{p} \\
x_\mathrm{p}z_\mathrm{p} &	  x_\mathrm{p}^2z_\mathrm{p} & x_\mathrm{p}y_\mathrm{p}z_\mathrm{p} & x_\mathrm{p}z_\mathrm{p}^2 & x_\mathrm{p}^2y_\mathrm{p}z_\mathrm{p} & (x_\mathrm{p}z_\mathrm{p})^2 & x_\mathrm{p}y_\mathrm{p}z_\mathrm{p}^2 & x_\mathrm{p}z_\mathrm{p}Z_\mathrm{p} & x_\mathrm{p}z_\mathrm{p}M_\mathrm{p} \\
y_\mathrm{p}z_\mathrm{p} &	  x_\mathrm{p}y_\mathrm{p}z_\mathrm{p} & y_\mathrm{p}^2z_\mathrm{p} & y_\mathrm{p}z_\mathrm{p}^2 & x_\mathrm{p}y_\mathrm{p}^2z_\mathrm{p} & x_\mathrm{p}y_\mathrm{p}z_\mathrm{p}^2 & (y_\mathrm{p}z_\mathrm{p})^2 & y_\mathrm{p}z_\mathrm{p}Z_\mathrm{p} & y_\mathrm{p}z_\mathrm{p}M_\mathrm{p} \\
Z_\mathrm{p} &	  x_\mathrm{p}Z_\mathrm{p} & y_\mathrm{p}Z_\mathrm{p} & z_\mathrm{p}Z_\mathrm{p} & x_\mathrm{p}y_\mathrm{p}Z_\mathrm{p} & x_\mathrm{p}z_\mathrm{p} & y_\mathrm{p}z_\mathrm{p}Z_\mathrm{p} & Z_\mathrm{p}^2 & M_\mathrm{p}Z_\mathrm{p} \\
M_\mathrm{p} &	  x_\mathrm{p}M_\mathrm{p} & y_\mathrm{p}M_\mathrm{p} & z_\mathrm{p}M_\mathrm{p} & x_\mathrm{p}y_\mathrm{p}M_\mathrm{p} & x_\mathrm{p}z_\mathrm{p}M_\mathrm{p} & y_\mathrm{p}z_\mathrm{p}M_\mathrm{p} & M_\mathrm{p}Z_\mathrm{p} & M_\mathrm{p}^2
	\end{bmatrix}\label{eq:EmMatrix}
\end{equation}

\subsection{Some Special Cases}
In order to gain some insights about the \textit{modus operandi}, it would be useful to consider some special cases. First we analyze what happens when no pixel is masked. In that case it can be shown that the matrix in Eq. \eqref{eq:EmMatrix} becomes diagonal. This is a consequence of the fact that when no pixel is masked
\begin{equation}
    \sum_\text{all pixels }x^a_\mathrm{p}y^b_\mathrm{p}z^c_\mathrm{p}\label{eq:TheQuantity}=0
\end{equation}
when either $a$, $b$ or $c$ are odd\footnote{This is a discrete version of the result
$$\iint_{S^2} x^ay^bz^c\ \mathrm{d}\Omega=(1+(-1)^a)(1+(-1)^b)(1+(-1)^c)\frac{((a-1)/2)!((b-1)/2)!((c-1)/2)!}{4((a+b+c+1)/2)!}$$ where the surface integral is performed over a unit sphere $S^2$. This is clearly zero when either $a$, $b$ or $c$ is odd.}. This further implies that when we have full sky information available then all the multipole parameters can be determined independently. But in reality, we always have a masked sky. In that case, the parameter extraction will depend upon where the series truncates. Thus in principle infinite number of parameters would be needed. But as we have seen that beyond quadrupole the power is small hence we truncated our series at $\ell=2$. We have also checked the effect of parameter extraction and $\ell$ value at which the series is terminated. These results are shown for monopole in Table \ref{tab:MonComp}.

Since the method is dependent upon information in unmasked pixels, we can in principle extract information in an extreme case when only one pixel information is available. But in reality we found that the method at \texttt{NSide=64} becomes unstable when we have information available in $\lesssim 10$ pixels. Additionally, we have also checked that the method works extremely well and calculates same values of the fitting parameters when we randomly mask the pixels. 
\subsection{Error Estimation \label{sec:ErrorSimu}}
\begin{figure}[t!]
	\centering
	\begin{tikzpicture}[node distance=1.5cm,
		every node/.style={fill=white, font=\sffamily}, align=center]
		% Specification of nodes (position, etc.)
		\node (block1)             [process]              {Extract Multipoles from\\ data map using Eq. \eqref{eq:Fitter}};
		\node (block2)     [process, below of=block1, yshift=-1cm]          {Prepare full sky maps\\ using these multipoles};
		\node (block3) [process, right of=block2, xshift=3.5cm] { Pixel Error\\  \texttt{Poisson}($N_\mathrm{p}$)};
		\node (block4) [process, right of=block3, xshift=3.5cm] {Extract multipole  \\  components};
		\node (block5) [process, above of=block4, yshift=1cm] {Obtain multipole \\ distribution, $\mu$ \& $\sigma$};
		%\node (block6) [process, above of=block3, yshift=1cm] {Prepare $I\in\{N,\alpha,B\}$\\Map (See Table \ref{tab:Observables})};
		\draw[->] (block1) -- (block2);
		\draw[->] (block2) -- (block3);
		\draw[->] (block3) -- (block4);
		\draw[->] (block4) -- (block5);
		%\draw[->] (block6) -- (block1);
		
		\draw [color=black,thick](3,-3.8) rectangle (12,-1.2);
		\node at (4,-4.3) [above=5mm, right=0mm] {{Repeat 10,000 times (Simulated Maps)}};
		
	\end{tikzpicture}
	\caption{Algorithm for generating mock maps and multipole extraction for number counts $N$. %The same method applies for extracting the multipole components from number counts $N$, {mean} spectral index $\alpha$ and mean {flux density} $B$.
	For details, the reader is referred to \S\ref{sec:ErrorSimu}.}
	\label{fig:Algorithm}
\end{figure}
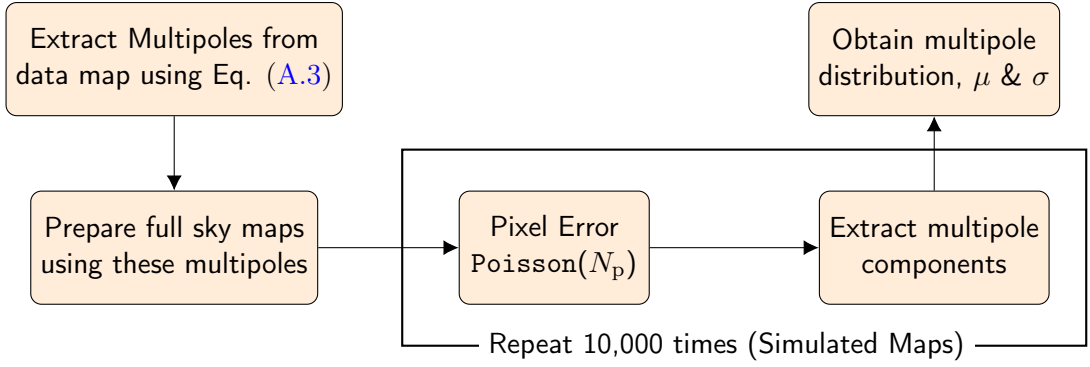

In order to estimate the errors in the fitted parameters, we perform simulations using the following algorithm, summarized in Figure \ref{fig:Algorithm}. 
\begin{enumerate}
\item \textit{Full Sky Map:} The monopole, dipole and quadrupole components from the data map are extracted using Eq. \eqref{eq:Fitter}. These multipoles are then used to prepare a full sky map $\mathbb{F}$ at \texttt{NSIDE} 64

\item \textit{Simulated Maps:} To incorporate error, we add, to the data value $N_\mathrm{p}$ of pixel p, a random number generated from the Poisson distribution \texttt{} using \texttt{numpy.random.poisson$(N_\mathrm{p})$}. This gives a simulated map $\mathbb{S}_j$. From this we again extract multipoles using Eq. \eqref{eq:Fitter}.% After this, we construct simulated maps,   . We fill the unmasked pixel $i$ of $\mathbb{F}$ with a random number generated  and add data value from the full sky map \item We extract multipole values from the simulated map $\mathbb{S}_j$
\item The previous step is repeated 10,000 times which gives a distribution for various extracted multipoles. The distribution, which is found to be Gaussian, is given in Figure \ref{fig:GaussFitNumCounts}. From this distribution, we can calculate relevant quantities of interest -- $\mu$ and $\sigma$. 
%\item From these values, the mean gives an estimate of the quantity under consideration and standard deviation gives an estimation of error.
\end{enumerate}

\bibliographystyle{JHEP}
\bibliography{main}

\end{document}